\documentclass[10pt,twocolumn,twoside,]{IEEEtran}
\usepackage{ntheorem}
\usepackage{amsmath}
\usepackage{amsfonts}
\newtheorem{theorem}{Theorem}
\newtheorem{corollary}{Corollary}
\newtheorem{lemma}{Lemma}
\newtheorem{definition}{Definition}
\newtheorem{proposition}{Proposition}

\newtheorem{remark}{Remark}
\usepackage{ntheorem}
\usepackage{amsmath}
\usepackage{amsfonts}
\usepackage{listings}
\usepackage{color}
\usepackage{array}
\usepackage{tabu}
\usepackage{mathrsfs,amssymb}
\usepackage{graphicx}
\graphicspath{ {figure/} }
\usepackage{float}
\usepackage{algorithm}

{}

\usepackage{algpseudocode}
\algdef{SE}[DOWHILE]{Do}{doWhile}{\algorithmicdo}[1]{\algorithmicwhile\ #1}%
\DeclareMathOperator*{\diag}{diag}
\begin{document}
	
	\title{Attack-Resilient Design for Connected and Automated Vehicles}
	
	%
	%
	%
	
\author{Tianci Yang \textsuperscript{1}, Carlos Murguia \textsuperscript{2}, Dragan Ne\v{s}i\'{c}\textsuperscript{3}, \emph{Fellow, IEEE}, and Chau Yuen\textsuperscript{1}, \emph{Fellow, IEEE}
	\thanks{\textsuperscript{1} T. Yang and Chau Yuen are with Nanyang Technological University, Singapore. Emails:
		\{tianci.yang, 	
		chau.yuen\}@ntu.edu.sg}%
	\thanks{\textsuperscript{2} Carlos Murguia is with the Department of Mechanical Engineering, Eindhoven University of Technology, The Netherlands. Email:
		c.g.murguia@tue.nl}
		\thanks{\textsuperscript{3} Dragan Ne\v{s}i\'{c} is with the Department of Electrical and Electronics Engineering, University of Melbourne, Australia. Email:
		dnesic@unimelb.edu.au}
}
	
	%
	%

\markboth{Journal of \LaTeX\ Class Files,~Vol.~14, No.~8, April~2020}%
{Shell \MakeLowercase{\textit{et al.}}: Bare Demo of IEEEtran.cls for IEEE Journals}
%



\maketitle

\begin{abstract}
	By sharing local sensor information via Vehicle-to-Vehicle (V2V) wireless communication networks, Cooperative Adaptive Cruise Control (CACC) is a technology that enables Connected and Automated Vehicles (CAVs) to drive autonomously on the highway in closely-coupled platoons. The use of CACC technologies increases safety and the traffic throughput, and decreases fuel consumption and CO2 emissions. However, CAVs heavily rely on embedded software, hardware, and communication networks that make them vulnerable to a range of cyberattacks. Cyberattacks to a particular CAV compromise the entire platoon as CACC schemes propagate corrupted data to neighboring vehicles potentially leading to traffic delays and collisions. Physics-based monitors can be used to detect the presence of False Data Injection (FDI) attacks to CAV sensors; however, unavoidable system disturbances and modelling uncertainty often translates to conservative detection results. Given enough system knowledge, adversaries are still able to launch a range of attacks that can surpass the detection scheme by hiding within the system disturbances and uncertainty -- we refer to this class of attacks as \textit{stealthy FDI attacks}. Stealthy attacks are hard to deal with as they affect the platoon dynamics without being noticed. In this manuscript, we propose a co-design methodology (built around a series convex programs) to synthesize distributed attack monitors and $H_{\infty}$ CACC controllers that minimize the joint effect of stealthy FDI attacks and system disturbances on the platoon dynamics while guaranteeing a prescribed platooning performance (in terms of tracking and string stability). Computer simulations are provided to illustrate the performance of out tools.
\end{abstract}


\section{Introduction}
Nowadays, our society is facing serious traffic congestion problems caused by the population increase in urban areas and limited road capacity. Taking advantage of recent sensing and wireless communication technologies, Connected and Automated Vehicles (CAVs) use on-board sensors and software to perceive the surroundings so that different driving tasks (e.g., lane keeping, vehicle following, and obstacle avoidance) can be carried out. Therefore, accurate sensor readings play a vital role for guaranteeing safety and security of CAVs. Inaccurate sensor information resulting from faults or cyberattacks can cause traffic accidents and even fatalities in the worst case. However, CAVs sensors remain vulnerable to adversarial attacks as they can be accessed remotely through compromised networks, road units, and vehicle computers. In \cite{petit2014potential}, Petit et al. show that radars and LiDARs that measure relative distance/velocity can be attacked remotely. In \cite{harris2015researcher}, the authors show that adversaries are able to compromise the LiDAR of a self-driving vehicle by generating artificial echoes of objects and vehicles. In \cite{yan2016can}, the vulnerability of millimeter-wave radars, ultrasonic sensors, and forward-looking cameras to attacks is investigated. They show that off-the-shelf hardware can be used to perform jamming and spoofing attacks on the Tesla Model S automobile. Moreover, in the context of Cooperative Adaptive Cruise Control (CACC), groups of CAVs drive together in tightly coupled platoons by sharing local sensor information with adjacent vehicles and roadside units via wireless communication networks \cite{lu2014connected}-\nocite{litman2017autonomous}\cite{anderson2014autonomous}. Therefore, for this class of cooperative driving schemes, the effect of malicious sensor readings in one vehicle propagates throughout the platoon leading to an amplified effect in the overall platoon dynamics. The above mentioned work highlights the fact that emerging CAVs not only bring convenience, efficiency, and comfort, but also many new security challenges to be addressed before these technologies can be widely used in society.

Existing results to deal with compromised sensors mainly focus on the problem of attack detection and isolation for vehicles under sensor attacks. For instance, in \cite{Liu2019,Wyk2019}, the authors exploit sensor redundancy and provide detection and isolation algorithms for a single vehicle under sensor attacks.
In \cite{mousavinejad2019distributed}, the authors provide an algorithm for detecting sensor attacks on connected vehicles using a set-membership filtering technique. In \cite{ju2020deception}, the problem of attack detection and estimation for connected vehicles under sensor attacks is solved using an unbiased finite impulse response (UFIR) estimator. In our previous work \cite{9502914}, we provide an attack-resilient sensor fusion algorithm for obtaining a secure estimation of vehicle states when some of the sensors are under attack. Then, the state estimation is used for attack detection and isolation.

Standard residual-based fault/attack detectors are widely used for identifying system faults and cyberattacks \cite{frank1997survey}\cite{isermann2005model}. The detection schemes compute the residual as the deviation between the sensor readings and the estimated values provided by a system estimator. If the residual crosses a predefined threshold, alarms are triggered. This standard detection procedure has also been shown effective for detecting faults/attacks on CAVs \cite{lopes2020active}-\nocite{wang2020anomaly}\nocite{yang2023robust}\cite{jeong2020sensor}. When anomalous sensors are detected, a CAV might stop the engine for preventing severe consequences, e.g., vehicle collisions. However, detection schemes are often conservative due to modeling uncertainties, sensor noise, network effects, etc. In many cases, attack signals can hardly be distinguished from the system disturbances under normal conditions. Therefore, by carefully designing the attack strategy, malicious signals that go undetected by the system attack detector can be injected into CAV sensors. We refer to this class of attacks as \textit{stealthy attacks}. Stealthy attacks can be a great threat to CAVs security since they affect the vehicle dynamics without being noticed. It follows that methodologies for constraining the capability of adversaries to inject stealthy attacks are highly needed.

In this manuscript, we consider platoons of CAVs driving cooperatively via CACC. We derive mathematical tools, in terms of semi-definite programs, for designing optimal fault/attack detectors and CACC controllers so that the platoon performance degradation caused by stealthy sensor attacks is minimal and not harmful. As a metric to quantify the impact of attacks, we use the ``size'' (in terms of volume) of the set of state trajectories (e.g., relative positions, velocities, and accelerations) that stealthy sensor attacks can induced in each CAV. We refer to these sets as stealthy reachable sets. Large (small) reachable sets imply that the attack can induce a large (small) set of driving trajectories in the vehicle. Even for linear models of CAVs, computing reachable sets exactly is computationally expensive and often not even tractable. Instead, we work with ellipsoidal outer approximations of reachable sets and design detectors and controllers to minimize their volume (reducing thus the size of the actual reachable set). As a second security metric, we use the minimum distance of stealthy reachable sets to critical states -- states that if reached compromise the integrity and safety of the vehicle, e.g., abnormally high/low velocities or too small inter-vehicle distances. This distance characterizes if stealthy attacks can actually cause a critical event or not given how they access the platoon dynamics (i.e., what sensors/vehicles are compromised). Hence, as design specification, we aim to synthesise detectors and controllers that make it impossible for stealthy attacks to reach a critical state.

There are a few related results in the literature.  In \cite{dadras2018reachable}, Dadras et al. provide a reachability analysis for a non-cooperative vehicle platoon, among which one malicious vehicle modifies its own acceleration randomly to degrade the platooning performance. Similarly, in \cite{sun2020impacts}, reachability analysis is provided for CAVs under random false-data injection and message delay attacks on V2V networks. We remark that the problem we consider is fundamentally different from the ones considered in \cite{dadras2018reachable} and \cite{sun2020impacts}. First, the injected false data in \cite{dadras2018reachable} and \cite{sun2020impacts} is \textit{randomly} constructed without using model knowledge. These attack signals can usually be detected by standard fault detectors and hence are, by definition, not stealthy. Besides, \cite{dadras2018reachable} and \cite{sun2020impacts} assume no system disturbances exist, even though noise in sensors and V2V communication networks and modeling uncertainty are unavoidable. Most importantly, \cite{dadras2018reachable} and \cite{sun2020impacts} focus on analysis, i.e., they analyze the attacker's reachable set after different system components such as estimators and controllers are already designed without security considerations. On the contrary, our work focuses on synthesis, i.e., system components are designed to minimize the impact of stealthy attacks quantified by our two proposed security metrics (stealthy reachable sets and distance to critical states). The mathematical machinery used here is inspired by the results in \cite{murguia2020security}, where reachability analysis approaches are provided for general Cyber-Physical Systems (CPSs) under different classes of disturbances and attacks. Summarizing, the main contribution of this manuscript is a set of synthesis tools (in terms of the solution a series of convex programs) to design robust state estimators, estimator-based fault/attack detectors, and CACC controllers for CAVs driving in platoons such that: 1) prescribed attack-free platooning performance is guaranteed (in terms of tracking and string stability); 2) the platooning performance degradation caused by stealthy sensors attacks is minimal (in terms of the size of stealthy reachable sets) and unharmful (characterized by strictly positive distances from reachable sets to critical states).

The paper is organized as follows. In Section \ref{pre}, some preliminary results needed for the subsequent sections are introduced. In Section \ref{sysd}, the considered vehicle platoon system is described. In Section \ref{design}, we provide tools for designing state estimators, an fault/attack detector, and controllers that minimize the attacker's capabilities with attack-free performance guaranteed. A method for evaluating the resilience of CAVs to stealthy attacks is given in Section \ref{critical}. Simulation study is conducted in Section \ref{simulation} to demonstrate the performance of our tools. Finally, in Section \ref{conclusion}, concluding remarks are given.

\section{Preliminaries}\label{pre}
\subsection{Notation}
We denote the set of real numbers by $\mathbb{R}$, the set of natural numbers by $\mathbb{N}$, and $\mathbb{R}^{n\times m}$ the set of $n\times m$ matrices for any $m,n \in \mathbb{N}$. For any vector $v\in\mathbb{R}^{n}$,  we denote {$v^{J}$} the stacking of all $v_{i}$, $i\in J$, $J\subset \left\lbrace 1,\hdots,n\right\rbrace$, $|v|=\sqrt{v^{\top} v}$. We say that a signal {$v(t):\mathbb{R}_{\geq 0}\to\mathbb{R}^{n}$} belongs to $l_{\infty}$, $\left\lbrace v(t)\right\rbrace \in l_{\infty}$, if $||v||_{\infty} <\infty$. $||v(t)||_{\mathcal{L}_{p}}$ is the p-norm of signal $v(t)$. We denote a variable $m$ uniformly distributed in the interval $(z_{1},z_{2})$ as $m\sim\mathcal{U}(z_{1},z_{2})$. 
\subsection{Definitions and Preliminary Results}
\begin{definition}[Vehicle String Stability]\emph{\cite{Ploeg2014}}\label{d1}
	Consider a string of $m\in N$ interconnected vehicles. This system is string stable
	if and only if
	\begin{equation}
		\begin{split}
			||z_{i}(t)||_{\mathcal{L}_{p}} \leq ||z_{i-1}(t)||_{\mathcal{L}_{p}},\hspace{2mm} \forall \hspace{1mm} t \geq 0, \hspace{1mm} 2 \leq i \leq m,	
		\end{split}
	\end{equation}
	where $z_{i}(t)$ can either be the intervehicle distance error $e_{i}(t)$, the velocity
	$v_{i}(t)$, or the acceleration $a_{i}(t)$ of vehicle i; $z_{1}(t)\in\mathcal{L}_{p}$ is a
	given input signal, and $z_{i}(0) = 0$ for $2 \leq i \leq m$.
\end{definition}
\begin{definition}[Reachable Set]\emph{\cite{murguia2020security}}
	Consider the perturbed Linear Time-Invariant (LTI) system:\label{def1}
	\begin{equation}\label{s1}
		\zeta(k+1)=\mathcal{A}\zeta(k)+\sum_{i=1}^{N}\mathcal{B}_{i}w_{i}(k),
	\end{equation}
	with $k \in \mathbb{N}$, state $\zeta\in\mathbb{R}^{n_{\zeta}}$, perturbation $w_{i}\in\mathbb{R}^{p_{i}}$ satisfying $w_{i}^{\top}W_{i}w_{i}\leq 1$ for some positive definite matrix $W_{i}\in\mathbb{R}^{p_{i}\times p_{i}}$, $i=1,\ldots,N, N\in\mathbb{N}$, and matrices $\mathcal{A}\in\mathbb{R}^{n_{\zeta}\times n_{\zeta}}$ and $\mathcal{B}_{i}\in\mathbb{R}^{n_{\zeta}\times p_{i}}$. The reachable set $\mathcal{R}_{k}^{\zeta}$ at time instant $k\geq0$ from the initial condition $\zeta_{1} \in \mathbb{R}^{n_{\zeta}}$ is the set of states reachable in $k$ steps by system \eqref{s1} through all possible disturbances satisfying $w_{i}(k)^{\top}W_{i}w_{i}(k)\leq 1$, i.e.,

	\begin{equation}
		\mathcal{R}^{\zeta}_{k} := \left\{ \zeta\in\mathbb{R}^{n_{\zeta}} \left|
		\begin{split}
			& \zeta = \zeta(k), \hspace{.5mm}\text{$\zeta(k)$ solution to \eqref{s1}},\\ & \text{and}\hspace{1mm} w_{i}(k)^{\top}W_{i}w_{i}(k)\leq 1,\\
		\end{split}
		\right. \right\}.
	\end{equation}

\end{definition}

\begin{corollary}[Ellipsoidal Approximation]\emph{\cite{murguia2020security}}\label{cor1}
	Consider the perturbed LTI system \eqref{s1} and the reachable set $\mathcal{R}_{k}^{\zeta}$ introduced in Definition \ref{def1}. For a given $a\in(0,1)$, if there exist constants $a_{1}$, $\ldots$, $a_{N}$ and matrix $P$ satisfying
	\begin{equation}\label{lmi1}
		\left\{\begin{split}
			&\text{\emph{s.t.}} \hspace{1mm}a_{1},\ldots, a_{N}\in(0,1), \hspace{.5mm} a_{1}+\ldots+ a_{N}\geq a,\\
			&P>0, \begin{bmatrix}
				aP&\mathcal{A}^{\top}P&\mathbf{0}\\
				P\mathcal{A}&P&P\mathcal{B}\\
				\mathbf{0}&\mathcal{B}^{\top}P&W_{a}
			\end{bmatrix}\geq 0,
		\end{split}\right.
	\end{equation}
	with $W_{a}:=\diag\begin{bmatrix}
		(1-a_{1})W_{1},\ldots,(1-a_{N})W_{N}
	\end{bmatrix}\in\mathbb{R}^{\bar{p}\times\bar{p}}$, $\mathcal{B}:=(\mathcal{B}_{1},\ldots,\mathcal{B}_{N})\in\mathbb{R}^{n_{\zeta}\times \bar{p}}$, and $\bar{p}=\sum_{i=1}^{N}p_{i}$; then, for all $k\geq k^{*}$, $\mathcal{R}_{k}^{\zeta}\subseteq \mathcal{E}_{k}^{\zeta}:=\left\lbrace \zeta^{\top}P^{\zeta}\zeta\leq\alpha_{k}^{\zeta}\right\rbrace $, with $\alpha_{k}^{\zeta}:=a^{k-1}\zeta(k^{*})^{\top}P\zeta(k^{*})+((N-a)(1-a^{k-1}))/(1-a)$.
\end{corollary}
\begin{lemma}[Projection]\emph{\cite{murguia2020security}}\label{lemma1}
	Consider the ellipsoid:
	\begin{equation}
		\begin{split}
			\mathcal{E}:=\left\lbrace x\in\mathbb{R}^{n}, y\in\mathbb{R}^{m}\bigg|\begin{bmatrix}
				x\\y
			\end{bmatrix}^{\top}\begin{bmatrix}
				Q_{1}&Q_{2}\\Q_{2}^{\top}&Q_{3}
			\end{bmatrix}\begin{bmatrix}
				x\\y
			\end{bmatrix}=\alpha \right\rbrace ,
		\end{split}
	\end{equation}
	for some positive definite matrix $Q\in\mathbb{R}^{(n+m)\times(n+m)}$ and constant $\alpha\in\mathbb{R}_{>0}$. The projection $\mathcal{E}'$ of $\mathcal{E}$ onto the $x$-hyperplane is given by the ellipsoid:
	\begin{equation}
		\mathcal{E}':=\left\lbrace x\in\mathbb{R}^{n}\vline \hspace{2mm}x^{\top}\begin{bmatrix}
			Q_{1}-Q_{2}Q_{3}^{-1}Q_{2}^{\top}
		\end{bmatrix}x=\alpha\right\rbrace .
	\end{equation}
\end{lemma}
\begin{figure}[t]\centering
	\includegraphics[width=0.5\textwidth]{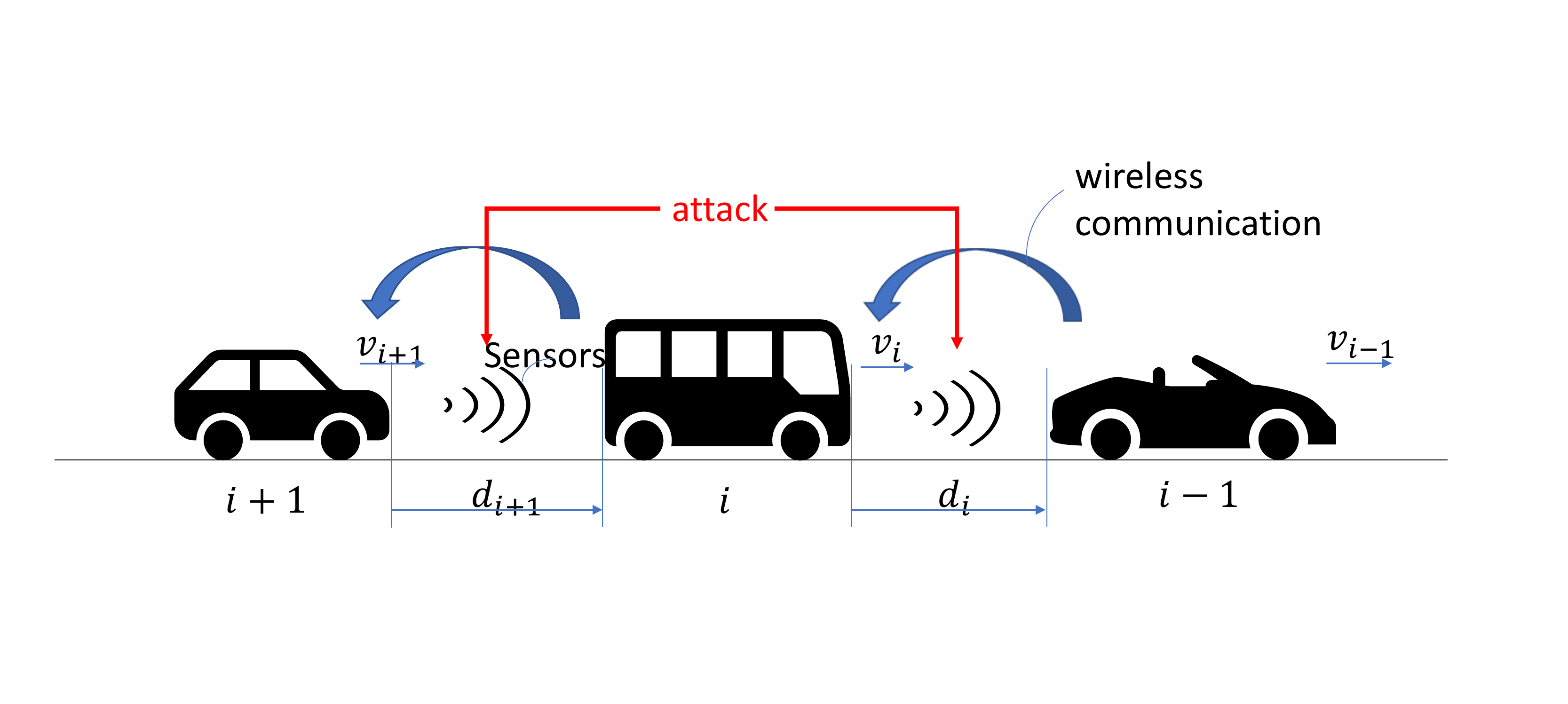}
	\caption{CACC-equipped vehicle platoon: each vehicle is equipped with radars which might be under attacks.}
	\centering
	\label{fig:1}
\end{figure} 
\section{System Description}\label{sysd}
In this section, we introduce the platoon dynamics that we consider, the control scheme, state estimator, and the fault/attack detector {for} CAVs.
\subsection{Vehicle Platoon Dynamics and the Control Scheme}
Consider a homogeneous string of $m$ vehicles as shown in Figure \ref{fig:1}. The distance between vehicle $i$ and its preceding vehicle $i-1$ is denoted as $d_{i}$, and its velocity is denoted as $v_{i}$. The objective of each vehicle is to keep a desired distance with its preceding vehicle:
\begin{equation}\label{distance}
	d_{r,i}(t)=s_{i}+hv_{i}(t), i\in S_{m},
\end{equation} 
with the time headway $h>0$, the standstill distance $s_{i}>0$. $S_{m}=\left\lbrace i\in\mathbb{N}|1\leq i\leq m\right\rbrace $ denotes the set of all vehicles in a platoon of length $m\in\mathbb{N}$. The spacing policy adopted here is expected to improve string stability  \cite{Paper1996, Ploeg2014}. The spacing error $e_{ri}(t)$ is then defined as 
\begin{equation}\label{e}
	\begin{split}
		e_{ri}(t)=&d_{i}(t)-d_{r,i}(t),\\
		=&(q_{i-1}(t)-q_{i}(t)-L_{i})-(s_{i}+hv_{i}(t)),
	\end{split}
\end{equation}
where $q_{i}$ denotes the rear-bumper position of vehicle $i$ and $L_{i}$ denotes its length. We consider the following vehicle model adopted in \cite{Ploeg2014}, 
\begin{equation}\label{vehicle}
	\begin{split}
		\begin{bmatrix}
			\dot{d}_{i}\\
			\dot{v}_{i}\\
			\dot{a}_{i}
		\end{bmatrix}=\begin{bmatrix}
			v_{i-1}-v_{i}\\
			a_{i}\\
			-\frac{1}{\tau}a_{i}+\frac{1}{\tau}u_{i}
		\end{bmatrix},\hspace{2mm}i\in S_{m},
	\end{split}
\end{equation}
with $\tau$ being a time constant representing
driveline dynamics, $a_{i}$ denoting the acceleration of vehicle $i$; $u_{i}$ denoting the desired accelerations of vehicles $i$. The controller described in \cite{Ploeg2014} that fulfills the vehicle-following objective and the string stability is adopted, where a new input $\epsilon_{i}$ is defined such that
\begin{equation}\label{ui}
	\begin{split}
		h\dot{u}_{i}=-u_{i}+\epsilon_{i},
	\end{split}
\end{equation}
with $\epsilon_{i}$ given as follows:
\begin{equation}\label{control}
	\begin{split}
		\epsilon_{i}=&K\begin{bmatrix}
			e_{ri}+\omega_{di1}+\delta_{i1}\\
			\dot{e}_{ri}+\omega_{di2}+\delta_{i2}
		\end{bmatrix} +u_{i-1}+\omega_{ui},\\
		=&K\left( \begin{bmatrix}
			e_{ri}\\
			\dot{e}_{ri}
		\end{bmatrix}+\omega_{di}+\delta_{i}\right)  +u_{i-1}+\omega_{ui},\hspace{1mm}i\in S_{m}.
	\end{split}
\end{equation}
where $K=\begin{bmatrix}
	k_{p}&k_{d}
\end{bmatrix}$. In \cite{Ploeg2014}, it has been proved that in the attack-free case, i.e., $\delta_{i}(t)=\mathbf{0}$, the vehicle following and string stability objectives are both fulfilled with any choice for
$k_{p},k_{d}>0$, $k_{d}>k_{p}\tau$.
The feedforward term $u_{i-1}$ is transmitted from vehicle $i-1$ to vehicle $i$ through a wireless communication channel, which is perturbed by channel noise. $\omega_{di}$ represent the noise in the radar sensor satisfying $\omega_{di}^{\top}\omega_{di}\leq\bar{\omega}_{1}$, $\omega_{ui}\in\mathbb{R}$ denotes the noise in the communication channel induced by network imperfections such as delay, pack dropouts etc., satisfying $\omega_{ui}^2\leq\bar{\omega}_{2}$. $\delta_{i}\in\mathbb{R}^{2}$ represents the injected attack signal into radars, i.e., $\delta_{i}(t)\neq \mathbf{0}$ for some $t\geq 0$ if the radars are compromised; otherwise $\delta_{i}(t)=\mathbf{0}$ for all $t\geq 0$. Let $\triangle v_{i}:=v_{i-1}-v_{i}$. Note that $\delta_{i}$ is a two-dimension vector, and this is because attacks on radars/lidars will affect $e_{ri}$ and $\dot{e}_{ri}$ both by modifying the information of $d_{i}$ and $\triangle v_{i}$.

Using the model \eqref{distance}-\eqref{control} with $x_{i}:=\begin{bmatrix}
	e_{ri}&v_{i}&a_{i}&\triangle v_{i}&a_{i-1}
\end{bmatrix}$, the following platoon model is obtained:
\begin{equation}\label{ee1}
	\begin{split}
		\begin{bmatrix}
			\dot{e}_{ri}\\
			\dot{v}_{i}\\
			\dot{a}_{i}\\
			\dot{\triangle v}_{i}\\
			\dot{a}_{i-1}
		\end{bmatrix}=&\begin{bmatrix}
			0&0&-h&1&0\\
			0&0&1&0&0\\
			0&0&-\frac{1}{\tau}&0&0\\
			0&0&-1&0&1\\
			0&0&0&0&-\frac{1}{\tau}
		\end{bmatrix}\begin{bmatrix}
			{e}_{ri}\\
			{v}_{i}\\
			{a}_{i}\\
			{\triangle v}_{i}\\
			a_{i-1}
		\end{bmatrix}\\
		&+\begin{bmatrix}
			0\\
			0\\
			\frac{1}{\tau}\\
			0\\
			0
		\end{bmatrix}u_{i}+\begin{bmatrix}
			0\\
			0\\
			0\\
			0\\
			\frac{1}{\tau}
		\end{bmatrix}u_{i-1}, \\
		=&A_{c}x_{i}+B_{c1}u_{i}+B_{c2}u_{i-1},\hspace{3mm} i\in S_{m}.
	\end{split}
\end{equation}
\eqref{control} can be rewritten as follows:
\begin{equation}\label{control2}
	\begin{split}
		\epsilon_{i}=&K\left( \begin{bmatrix}
			1&0&0&0&0\\
			0&0&-h&1&0
		\end{bmatrix}x_{i}+\omega_{di}+\delta_{i}\right) +u_{i-1}+\omega_{ui},\\
		=&K(Cx_{i}+w_{di}+\delta_{i})+u_{i-1}+\omega_{ui},\\
		=&Ky_{i}+u_{i-1}+\omega_{ui},\hspace{1mm}i\in S_{m}.
	\end{split}
\end{equation}
with $y_{i}\in\mathbb{R}^{2}$ and $C\in\mathbb{R}^{2\times 5}$ defined accordingly. Similarly, substituting \eqref{control2} into \eqref{ui} gives
\begin{equation}\label{u}
	\dot{u}_{i}=-\frac{1}{h}u_{i}+\frac{1}{h}(Ky_{i}+u_{i-1}+\omega_{ui}),\hspace{1mm}i\in S_{m}.
\end{equation}
The first vehicle, without a preceding vehicle in front, will follow a virtual reference vehicle $(i=0)$, so that the same controller as the other vehicles can be employed to the lead vehicle. We formulate the virtual reference vehicle as follows:
\begin{equation}\label{ee2}
	\begin{split}
		\begin{bmatrix}
			\dot{e}_{r0}\\
			\dot{v}_{0}\\
			\dot{a}_{0}\\
			\dot{u}_{0}
		\end{bmatrix}=\begin{bmatrix}
			0&0&0&0\\0&0&1&0\\0&0&-\frac{1}{\tau}&\frac{1}{\tau}\\0&0&0&-\frac{1}{h}
		\end{bmatrix}\begin{bmatrix}
			e_{r0}\\
			v_{0}\\
			a_{0}\\
			u_{0}
		\end{bmatrix}+\begin{bmatrix}
			0\\0\\0\\-\frac{1}{h}
		\end{bmatrix}\epsilon_{0},
	\end{split}
\end{equation}
where $\epsilon_{0}$ denotes external platoon input.

The closed-loop system \eqref{ee1}-\eqref{u} can be written in a compact form as follows:
\begin{equation}\label{ss}
	\left\{\begin{split}
		\dot{x}_{i}=&A_{c}x_{i}+B_{c1}u_{i}+B_{c2}u_{i-1},\\
		y_{i}=&Cx_{i}+\omega_{di}+\delta_{i},\\
		\dot{u}_{i}=&-\frac{1}{h}u_{i}+\frac{1}{h}Ky_{i}+\frac{1}{h}u_{i-1}+\frac{1}{h}\omega_{ui},
	\end{split}\right.
\end{equation}
with 
\begin{equation}\label{matrix11}
	\begin{split}
		A_{c}=&\begin{bmatrix}
			0&0&-h&1&0\\
			0&0&1&0&0\\
			0&0&-\frac{1}{\tau}&0&0\\
			0&0&-1&0&1\\
			0&0&0&0&-\frac{1}{\tau}
		\end{bmatrix},B_{c1}=\begin{bmatrix}
			0\\
			0\\
			\frac{1}{\tau}\\
			0\\
			0
		\end{bmatrix}, B_{c2}=\begin{bmatrix}
			0\\
			0\\
			0\\
			0\\
			\frac{1}{\tau}
		\end{bmatrix}, \\
		C=&\begin{bmatrix}
			1&0&0&0&0\\
			0&0&-h&1&0
		\end{bmatrix}, K=\begin{bmatrix}
			k_{p}&k_{d}
		\end{bmatrix}.
	\end{split}
\end{equation}
Adopting exact discretization method with sampling time $T_{s}$ to discretize \eqref{ss},  
\begin{subequations}\label{cl}
	\begin{align}
		&x_{i}(k+1)=Ax_{i}(k)+B_{1}u_{i}(k)+B_{2}u_{i-1}(k),\label{ssd1}\\
		&y_{i}(k)=Cx_{i}(k)+\omega_{di}(k)+\delta_{i}(k),\label{ssd2}\\
		&{u}_{i}(k+1)=a_{u}u_{i}(k)+b_{u}Ky_{i}(k)+b_{u}u_{i-1}(k)+b_{u}\omega_{ui}(k),\label{ssd3}
	\end{align}
\end{subequations}
with
\begin{equation}\label{matrix12}
	\begin{split}
		A=&e^{A_{c}T_{s}},B_{1}=\left( \int_{0}^{T_{s}}e^{A_{c}(T_{s}-s)}ds\right) B_{c1},\\
		B_{2}=&\left( \int_{0}^{T_{s}}e^{A_{c}(T_{s}-s)}ds\right) B_{c2},
		a_{u}=e^{-\frac{1}{h}T_{s}}, \\
		b_{u}=&\left( \int_{0}^{T_{s}}e^{-\frac{1}{h}(T_{s}-s)}ds\right)\frac{1}{h}.
	\end{split}
\end{equation}
\subsection{Estimator and Residual}\label{estimation}
Here, we consider CAVs are equipped with radars, speed and acceleration sensors for measuring the relative distance, relative velocity, velocity and acceleration respectively. An estimator is utilized by each vehicle in the platoon to compute state estimates from noisy sensor measurements. The estimation residual is used then by a detector for identifying sensor attacks.

{Here, we provide an estimator with the following structure for system \eqref{ssd1}.}

\begin{equation}\label{sse}
	\begin{split}
		\hat{x}_{i}(k+1)=&A\hat{x}_{i}(k)+B_{1}u_{i}(k)+B_{2}(u_{i-1}(k)+\omega_{ui}(k))\\
		&+L\big(y_{ei}(k)-C_{e}\hat{x}_{i}(k)\big),
	\end{split}
\end{equation}
with $y_{ei}(k)=C_{e}x_{i}(k)+\omega_{ei}(k)+\Gamma\delta_{i}(k)$,
\begin{equation}
	C_{e}=\begin{bmatrix}
		1&0&0&0&0\\
		0&1&0&0&0\\
		0&0&1&0&0\\
		0&0&0&1&0
	\end{bmatrix}, \Gamma=\begin{bmatrix}
		1&0\\
		0&0\\
		0&0\\
		0&1
	\end{bmatrix},
\end{equation}
$\omega_{ei}\in\mathbb{R}^{4}$ denoting the sensor noise satisfying $\omega_{ei}^{\top}\omega_{ei}\leq\bar{\omega}_{3}$, $\hat{x}_{i}(k)\in\mathbb{R}^{5}$ is the state estimate of $x_{i}$ and $L\in\mathbb{R}^{5\times 4}$ is the estimator gain. It can be verified that $(A,C_{e})$ is observable. Define the estimation error $e_{i}:=x_{i}-\hat{x}_{i}$. Given the system dynamics \eqref{cl} and the estimator \eqref{sse}, the estimation error dynamics is described by the following difference equation:
\begin{equation}\label{ssee}
	\begin{split}
		e_{i}(k+1)=&(A-LC_{e})e_{i}(k)-B_{2}\omega_{ui}(k)\\
		&-L\omega_{ei}(k)-L\Gamma\delta_{i}(k).
	\end{split}
\end{equation}
\begin{remark}
	Since the error dynamics of the estimator designed for \eqref{ssd1} depends on $\delta_{i}$, it can be used for detecting faults or attacks on sensors. 
\end{remark}
	In the attack-free case, i.e., $\delta_{i}(k)=\mathbf{0}$ for $k\geq 0$, we have
	\begin{equation}\label{ssee1}
		\begin{split}
			\bar{e}_{i}(k+1)=&(A-LC_{e})\bar{e}_{i}(k)-B_{2}\omega_{ui}(k)-L\omega_{ei}(k),
		\end{split}
	\end{equation}
	where $\bar{e}_{i}$ represents the estimation error in the attack-free case, i.e., $\delta_{i}(k)=\mathbf{0}$ for $k\geq 0$.
		Define the residual as follows
		\begin{equation}\label{r1}
			\begin{split}
				r_{i}(k)=&y_{ei}(k)-C_{e}\hat{x}_{i}(k).
			\end{split}
		\end{equation}
		The estimation error and the residual evolves according to the difference equation:
		\begin{equation}\label{er}
			\left\{\begin{split}	e_{i}(k+1)=&(A-LC_{e})e_{i}(k)-B_{2}\omega_{ui}(k)\\
				&-L\omega_{ei}(k)-L\Gamma\delta_{i}(k),\\
				r_{i}(k)=&C_{e}e_{i}(k)+\omega_{ei}(k)+\Gamma\delta_{i}(k),
			\end{split}
			\right.
		\end{equation}
		and in the attack-free case, we have
		\begin{equation}\label{err}
			\left\{\begin{split}	\bar{e}_{i}(k+1)=&(A-LC_{e})e_{i}(k)-B_{2}\omega_{ui}(k)-L\omega_{ei}(k),\\
				r_{i}(k)=&C_{e}e_{i}(k)+\omega_{ei}(k).
			\end{split}
			\right.
		\end{equation}
		\subsection{System Monitor}
		Next, we develop a standard fault detector for each CAV in the platoon such that if estimation residual in \eqref{er} is larger than expected, alarms are triggered and a fault or an attack on the vehicle is detected. 
		Consider the detection procedure is a distance measure $z_{i}\in\mathbb{R}$ that measures how deviated the estimator is from the attack-free dynamics. A quadratic form of the residual is considered here, i.e., $z_{i}:=r_{i}^{\top}\Pi r_{i}$ for some positive definite matrix $\Pi\in\mathbb{R}^{4\times 4}$. Consider the monitor given in \eqref{monitor}.\\
		\begin{algorithm}[h]
			\textbf{System Monitor:}\\
			\begin{equation}\label{monitor}
				\text{If}\hspace{1mm}z_{i}(k)=r_{i}(k)^{\top}\Pi r_{i}(k)>1, \text{for some}\hspace{1mm}k.
			\end{equation}
			\textbf{Design parameter:} positive semidefinite matrix $\Pi\in\mathbb{R}^{4\times 4}$.\\
			\textbf{Output:} alarm time(s) $k$. 
		\end{algorithm}
		
		The monitor is designed such that alarms are triggered if $z_{i}(k)>1$. Therefore, $z_{i}(k)\leq 1$ is supposed to contain all attack-free trajectories, i.e., $\Pi$ is selected such that for all $k\geq k^{*}$, the ellipsoid $r_{i}(k)^{\top}\Pi r_{i}(k)\leq 1$ contains all the possible trajectories that $\omega_{ui}(k)$, $\omega_{ei}(k)$ can induce in the residual given in \eqref{er} with $\omega_{ui}(k)^{\top}\omega_{ui}(k)\leq\bar{\omega}_{2}$, $\omega_{ei}^{\top}(k)\omega_{ei}(k)\leq\bar{\omega}_{3}$, $\delta_{i}(k)=\mathbf{0}$. 
		
\begin{figure}[t]\centering
			\includegraphics[width=0.5\textwidth]{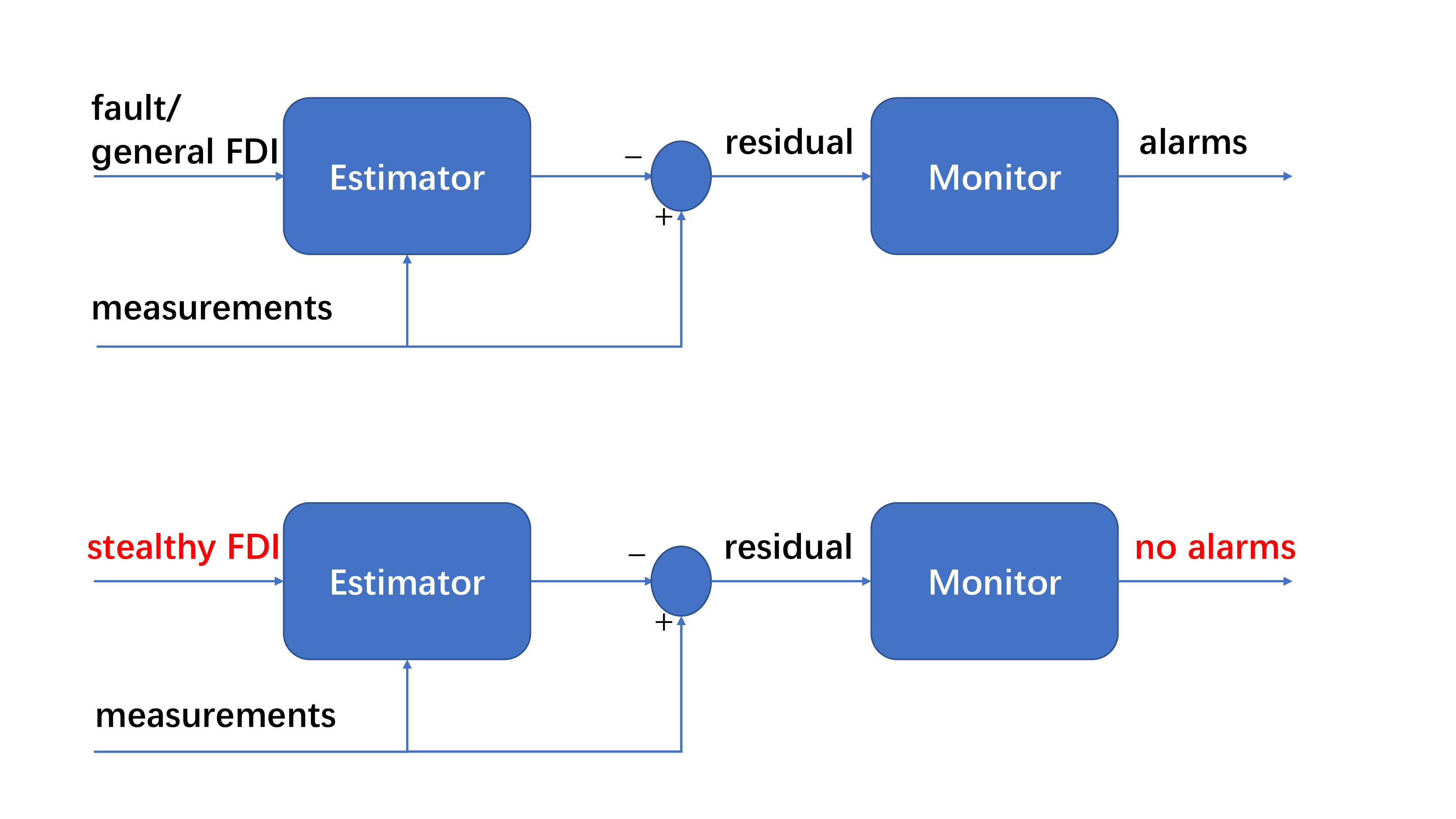}
			\caption{A standard fault/attack detector is useful for identifying a large class of anomalous behaviors including faults or general FDI attacks; but it cannot work in the presence of stealthy FDI attacks.}
			\centering
			\label{fig:x}
		\end{figure}

		Note that this is a novel fault/attack detector that is useful for identifying faults (sensor, actuator, and dynamics faults) and \emph{non-stealthy} FDI attacks, which covers a large class of anomalous behaviors. However, by definition, no monitor (including \eqref{monitor}) can identify stealthy FDI attacks. This is schematically depicted in Figure \ref{fig:x}.

			\section{Resilient Estimator, Detector, and Controller Design}\label{design}
			In this section, we assume the attacker has knowledge of the vehicle model and the detector structure so that the injected $\delta_{i}$ does not trigger alarms by monitor \eqref{monitor}. For each CAV in a platoon, we aim to design the estimator, the detector, and the controller that are resilient to stealthy radar attacks, i.e., the damage caused by stealthy attacks on radars is minimized. This class of stealthy attacks can be characterized as a constrained control problem in $\delta_{i}$:
			\begin{equation}\label{stealthy}
				\left\lbrace \delta_{i} \in \mathbb{R} \hspace{1mm} \vline \hspace{1mm} \text{$r_{i}(k)$ satisfies \eqref{er} and }r_{i}(k)^{\top}\Pi r_{i}(k)\leq 1\right\rbrace.
			\end{equation}
			The closed-loop systems \eqref{cl} can be written as follows:
			\begin{equation}\label{closed}
				\left\{\begin{split}
					x_{i}(k+1)=&Ax_{i}(k)+B_{1}u_{i}(k)+B_{2}u_{i-1}(k),\\
					{u}_{i}(k+1)=&a_{u}u_{i}(k)+b_{u}KCx_{i}(k)+b_{u}u_{i-1}(k)\\
					&+b_{u}\omega_{ui}(k)+b_{u}K\omega_{di}(k)+b_{u}K\delta_{i}(k).
				\end{split}
				\right.
			\end{equation}
			We are interested in the vehicle state trajectories that the attacker can induce by injecting attack signals satisfying \eqref{stealthy} into the radar sensors. This motivates the following definition.
			\begin{definition}[Stealthy Reachable Set]
				The stealthy reachable set of vehicle $i$ at time-instant $k$, $\mathcal{R}_{k}^{x_{i}}$, is the set of all states reachable by system \eqref{ssd1} through all possible initial conditions and disturbances satisfying $u_{i-1}^{2}\leq\bar{u}$, $\omega_{di}^{\top}(k)\omega_{di}(k)\leq\bar{\omega}_{1}$ $\omega_{ui}^{\top}\omega_{ui}\leq\bar{\omega}_{2}$, and attack sequences satisfying \eqref{stealthy}, i.e.,
				\begin{equation}\label{rx}
					\mathcal{R}_{k}^{x_{i}} := \left\{ x_{i} \in \mathbb{R}^{5} \left|
					\begin{split}
						& x_{i} \text{ and } u_{i} \hspace{1mm}\text{satisfy \eqref{closed2}},\\ & \delta_{i}\hspace{1mm}\text{satisfies}\hspace{1mm}\eqref{stealthy}, u_{i-1}^{2}\leq\bar{u},\\
						&\omega_{di}^{\top}(k)\omega_{di}(k)\leq\bar{\omega}_{1},\\ &\omega_{ui}(k)^{\top}\omega_{ui}(k)\leq\bar{\omega}_{2}, k\in\mathbb{N}.\\ 
					\end{split}
					\right. \right\}.
				\end{equation}
			\end{definition}
			
			Since $\mathcal{R}_{k}^{x_{i}}$ contains all the state {trajectories} that the attacker can induce by injecting stealthy attacks, we use the volume of the set $\mathcal{R}_{k}^{x_{i}}$ to quantify the attacker's capabilities \cite{murguia2020security}. Note that it is in general not tractable to compute $\mathcal{R}_{k}^{x_{i}}$ exactly, we look for an outer ellipsoidal approximation of the form $\mathcal{E}_{k}^{x_{i}} := \left\lbrace x_{i}\in\mathbb{R}^{5}|\hspace{1mm}x_{i}^{\top}P^{x_{i}}x_{i}\leq\alpha_{k}^{x_{i}}\right\rbrace $ such that $\mathcal{R}_{k}^{x_{i}}\subseteq\mathcal{E}_{k}^{x_{i}}$ for all $k \in \mathbb{N}$. This means that the ellipsoid $x_{i}^{\top}P^{x}x_{i}=\alpha_{k}^{x_{i}}$ contains all the possible trajectories induced by stealthy attacks satisfying \eqref{stealthy}. For Linear Time-Invariant (LTI) systems, $\mathcal{E}_{k}^{x_{i}}$ is a good approximation of $\mathcal{R}_{k}^{x_{i}}$ and it can be computed efficiently using LMIs, hence, we use the volume of $\mathcal{E}_{k}^{x_{i}}$ instead for assessing attacker's  capabilities.

			Consider $r_{i}(k)$ in \eqref{er}, since $\Gamma$ has full column rank, we can compute $\delta_{i}(k)$ as follows:
			\begin{equation}\label{delta}
				\begin{split}
					\delta_{i}(k)=\Gamma^{\dagger}\big(r_{i}(k)-C_{e}e_{i}(k)-\omega_{ei}(k)\big),
				\end{split}
			\end{equation}
			where $\Gamma^{\dagger}$ denotes the Moore-Penrose inverse of $\Gamma$.
			Substituting \eqref{delta} into \eqref{closed} yields
			\begin{equation}\label{closed2}
				\left\{\begin{split}
					x_{i}(k+1)=&Ax_{i}(k)+B_{1}u_{i}(k)+B_{2}u_{i-1}(k),\\
					{u}_{i}(k+1)=&a_{u}u_{i}(k)+b_{u}KCx_{i}(k)+b_{u}u_{i-1}(k)\\
					&+b_{u}\omega_{ui}(k)+b_{u}K\omega_{di}(k)\\
					&+b_{u}K\Gamma^{\dagger}(r_{i}(k)-C_{e}e_{i}(k)-\omega_{ei}(k)).
				\end{split}
				\right.
			\end{equation}
			Define the extended state $\zeta_{i}=\begin{bmatrix}
				x_{i}^{\top}&u_{i}^{\top}
			\end{bmatrix}^{\top}$, the closed-loop dynamics \eqref{closed} can be written compactly as follows:
			\begin{equation}\label{zeta}
				\begin{split}
					\zeta_{i}(k+1)=&\mathcal{A}\zeta_{i}(k)+\mathcal{B}_{1}u_{i-1}(k)+\mathcal{B}_{2}\omega_{di}(k)+\mathcal{B}_{3}\omega_{ui}(k)\\
					&+\mathcal{B}_{4}\omega_{ei}(k)+\mathcal{B}_{5}r_{i}(k)+\mathcal{B}_{6}e_{i}(k),
				\end{split}
			\end{equation}
			with
			\begin{equation}\label{matrices}
				\begin{split}
					\mathcal{A}=&\begin{bmatrix}
						A&B_{1}\\
						b_{u}KC&a_{u}
					\end{bmatrix},\\
					\mathcal{B}_{1}=&\begin{bmatrix}
						B_{2}\\
						b_{u}
					\end{bmatrix}, \mathcal{B}_{2}=\begin{bmatrix}
						\mathbf{0}\\
						b_{u}K\\
					\end{bmatrix},\\
					\mathcal{B}_{3}=&\begin{bmatrix}
						\mathbf{0}\\
						b_{u}
					\end{bmatrix},\mathcal{B}_{4}=\begin{bmatrix}
						\mathbf{0}\\
						-b_{u}K\Gamma^{\dagger}
					\end{bmatrix}, \\
					\mathcal{B}_{5}=&\begin{bmatrix}
						\mathbf{0}\\
						b_{u}K\Gamma^{\dagger}
					\end{bmatrix}, \mathcal{B}_{6}=\begin{bmatrix}
						\mathbf{0}\\
						-b_{u}K\Gamma^{\dagger}C_{e}
					\end{bmatrix}.
				\end{split}
			\end{equation}
			The closed-loop dynamics \eqref{zeta} is now written in the form \eqref{s1} driven by peak-bounded perturbations $u_{i-1}$, $\omega_{di}$, $\omega_{ei}$, $\omega_{ui}$, $r_{i}$, and $e_{i}$. The residual $r_{i}$ acts as a peak-bounded perturbation because the attack sequence $\delta_i$ satisfies \eqref{stealthy} (i.e., $r_i(k)^\top \Pi r_i(k) \leq 1$ for all $k \in \mathbb{N}$). Next, we introduce the method of selecting $L$ and $\Pi$ such that $e_{i}(k)^{\top}P^{e}e_{i}(k)\leq \alpha_{k}^{e_{i}}$ for some $P^{e}>0$ and $\alpha_{k}^{e_{i}}>0$.
			
			Define the reachable set for \eqref{zeta}:
			\begin{equation}\label{rzeta}
				\mathcal{R}_{k}^{\zeta_{i}} := \left\{ \zeta_{i} \in  \mathbb{R}^{6} \left|
				\begin{split}
					& \zeta_{i}\hspace{1mm}\text{satisfies \eqref{zeta}}, r_i(k)^\top \Pi r_i(k) \leq 1,\\
					&e_{i}(k)^{\top}P^{e}e_{i}(k)\leq \alpha_{k}^{e_{i}},\\
					&u_{i-1}(k)^{\top}u_{i-1}(k)\leq\bar{u},\\ &\omega_{di}(k)^{\top}\omega_{di}(k) \leq \bar{\omega}_{1},\\
					&\omega_{ui}(k)^{\top}\omega_{ui}(k) \leq \bar{\omega}_{2}, \\
					&\omega_{ei}(k)^{\top}\omega_{ei}(k) \leq \bar{\omega}_{3}, k \in \mathbb{N}.
				\end{split}
				\right. \right\}.
			\end{equation}
			\begin{remark}\label{remark2}
				We are ultimately interested in the stealthy reachable set of the vehicle states $\mathcal{R}_{k}^{x_{i}}$ introduced in \eqref{rx}. Note that $\mathcal{R}_{k}^{x_{i}}$ is the projection of $\mathcal{R}_{k}^{\zeta_{i}}$ onto the $x_{i}$-hyperplane. Hence, if $\mathcal{R}_{k}^{\zeta_{i}}\subseteq\mathcal{E}_{k}^{\zeta_{i}}$, then $\mathcal{R}_{k}^{x_{i}}\subseteq\mathcal{E}_{k}^{\zeta_{i}}||_{x_{i}}$, where $\mathcal{E}_{k}^{\zeta_{i}}||_{x_{i}}$ denotes the projection of $\mathcal{E}_{k}^{\zeta_{i}}$ onto the $x_{i}$-hyperplane. Therefore, to obtain the ellipsoid $\mathcal{E}_{k}^{x_{i}}$ containing $\mathcal{R}_{k}^{x_{i}}$, we can first obtain $\mathcal{E}_{k}^{\zeta_{i}}$ containing $\mathcal{R}_{k}^{x_{i}}$ and then {project it onto the $x_{i}$-hyperplane} to obtain $\mathcal{E}_{k}^{x_{i}}$.
			\end{remark}
			\subsection{Resilient Estimator and Monitor}
			From \eqref{rzeta}, it can be seen that if we select $L$ and $\Pi$ such that the volumes of ellipsoids $e_{i}(k)^{\top}P^{e}e_{i}(k)\leq \alpha_{k}^{e_{i}}$ and $r_{i}^{\top}(k)\Pi r_{i}(k)\leq 1$ are reduced, then the volume of $\mathcal{R}_{k}^{\zeta_{i}}$ will also be reduced. This indicates such estimator and monitor restrict attacker's capabilities and show higher resilience towards stealthy attacks.
			
			Substituting \eqref{delta} into \eqref{er} yields
			\begin{equation}\label{er1}
				\begin{split}
									e_{i}(k+1)
					=&(A-L(I-\Gamma\Gamma^{\dagger})C_{e})e_{i}(k)-B_{2}\omega_{ui}(k)\\
					&-L(I-\Gamma\Gamma^{\dagger})\omega_{ei}(k)-L\Gamma\Gamma^{\dagger}r_{i}(k).
				\end{split}
			\end{equation}
			The reachable set of the estimation error with dynamics \eqref{er1} is given as
			\begin{equation}\label{re}
				\mathcal{R}_{k}^{e_{i}} := \left\{ e_{i} \in \mathbb{R}^{5} \left|
				\begin{split}
					& e_{i} \hspace{1mm}\text{satisfy \eqref{er1}},\\ &r_i(k)^\top \Pi r_i(k) \leq 1,\\ 
					&\omega_{ui}(k)^{\top}\omega_{ui}(k)\leq\bar{\omega}_{2},\\ &\omega_{ei}(k)^{\top}\omega_{ei}(k)\leq\bar{\omega}_{3},
				\end{split}
				\right. \right\}.
			\end{equation}

				Since it is analytically intractable to compute $\mathcal{R}_{k}^{\zeta_{i}}$ and $\mathcal{R}_{k}^{e_{i}}$, we use Corollary 1 to derive outer ellipsoidal bounds of the form $\mathcal{E}_{k}^{\zeta_{i}}=\left\lbrace \zeta_{i}\in\mathbb{R}^{6}|\zeta_{i}(k)^{\top}P^{\zeta}\zeta_{i}(k)\leq\alpha_{k}^{\zeta_{i}}\right\rbrace $ and $\mathcal{E}_{k}^{e_{i}}=\left\lbrace e_{i}\in\mathbb{R}^{5}| e_{i}(k)^{\top}P^{e}e_{i}(k)\leq\alpha_{k}^{e_{i}}\right\rbrace$ such that $\mathcal{R}_{k}^{\zeta_{i}}\subseteq\mathcal{E}_{k}^{\zeta_{i}}$ and $\mathcal{R}_{i}^{e_{i}}\subseteq\mathcal{E}_{k}^{e_{i}}$. Note that the $e_{i}$ dynamics in \eqref{er1} does not depend on $\zeta_{i}(k)$. Thus we first compute $\mathcal{E}_{k}^{e_{i}}$ and then analyze how it propagates to $\mathcal{E}_{k}^{\zeta_{i}}$. 
				Note that the dynamics of $\zeta_{i}$ given in \eqref{zeta} depends on bounded disturbances $e_{i}$ and $r_{i}$ both. Therefore, we use Corollary 1 to select $L$ and $\Pi$ such that a weighted sum of the volumes of $\mathcal{E}_{k}^{e_{i}}$ and $r_{i}(k)^{\top}\Pi r_{i}(k)=1$ is minimized.
				\begin{theorem}\label{th1}
					Consider the estimation error dynamics \eqref{er1} with matrices $(A, C_{e}, B_{2}, \Gamma)$, the perturbation bounds $\bar{\omega}_{2}$, $\bar{\omega}_{3}$. For a given $a\in(0,1)$, tuning parameters $\alpha_{1},\alpha_{2}>0$ with $\alpha_{1}+\alpha_{2}=1$, $\epsilon>0$, if there exist constants $a_{1}$, $a_{2}$, $a_{3}$, and matrices $P^{e}$, $L$, $\tau_{1}$, $\tau_{2}$, $\Pi$ solution of the following optimization:
					\begin{equation}\label{lmi3}
						\left\{\begin{split}
							&\min_{P^{e},\Pi,a_{1},a_{2}, a_{3}}-\alpha_{1}\log\det[P^{e}]-\alpha_{2}\log\det[\Pi],\\
							&s.t. \hspace{1mm}a_{1},a_{2}, a_{3}\in(0,1), a_{1}+a_{2}+a_{3}\geq a,\\
							&\tau_{1}\geq 0, \tau_{2}\geq 0,P^{e}>0, {\Pi}>0,\\
							& L_{1}:=\begin{bmatrix}
								aP^{e}&A^{\top}P^{e}-C_{e}^{\top}(I-\Gamma\Gamma^{\dagger})^{\top}L^{\top}P^{e}&\mathbf{0}\\
								*&P^{e}&Z\\
								*&*&{W}_{a}
							\end{bmatrix}\geq 0,\\
							& L_{2}:=\begin{bmatrix}
								f_{1}&-C_{e}^{\top}\Pi&\mathbf{0}\\
								*&\tau_{2}I-\Pi&\mathbf{0}\\
								*&*&f_{2}
							\end{bmatrix}\geq 0,\\
							&f_{1}=\tau_{1}P^{e}-C_{e}^{\top}\Pi C_{e},\\
							&f_{2}=1-\tau_{1}(\alpha_{\infty}^{e_{i}}+\epsilon)-\tau_{2}\bar{\omega}_{3},
						\end{split}\right.
					\end{equation}
					with $Z=\begin{bmatrix}
						-P^{e}B_{2}&-P^{e}L(I-\Gamma\Gamma^{\dagger})&-P^{e}L\Gamma\Gamma^{\dagger}
					\end{bmatrix}$, ${W}_{a}=\diag\begin{bmatrix}
						\frac{1-a_{1}}{\bar{\omega}_{2}},\frac{1-a_{2}}{\bar{\omega}_{3}}I_{4}, (1-a_{3}){\Pi}
					\end{bmatrix}\in\mathbb{R}^{9\times 9}$, $\alpha_{\infty}^{e_{i}}=(3-a)/(1-a)$. Then for all $k\geq1$, $\mathcal{R}_{k}^{e_{i}}\subseteq\mathcal{E}_{k}^{e_{i}}:=\left\lbrace e_{i}\in\mathbb{R}^{5}|e_{i}^{\top}(k)P^{e}e_{i}(k)\leq\alpha_{k}^{e_{i}}\right\rbrace $, and $\alpha_{k}^{e_{i}}:=a^{k-1}e_{i}(1)^{\top}P^{e}e_{i}(1)+\big((3-a)(1-a^{k-1})\big)/(1-a)$, the monitor $r_{i}(k)^{\top}\Pi r_{i}(k)\leq 1$ is satisfied for all $e_{i}(k)$ satisfying $e_{i}(k)^{\top}P^{e}e_{i}(k)\leq\alpha_{\infty}^{e_{i}}+\epsilon$ and $\omega_{ei}(k)$ satisfying $\omega_{ei}(k)^{\top}\omega_{ei}(k)\leq\bar{\omega}_{3}$, and the weighted sum of the volumes of ellipsoid $\mathcal{E}_{k}^{e_{i}}$ and $r_{i}(k)^{\top}\Pi r_{i}(k)=1$ is minimized in the sense of Corollary 1.
				\end{theorem}
				\textit{Proof:}
				The reachable set $\mathcal{R}_{k}^{e_{i}}$ in \eqref{re} is a LTI system driven by peak-bounded perturbations. It follows that, under the conditions stated in Theorem \ref{th1}, Corollary 1 can be used to obtain outer ellipsoidal approximations of the form $\mathcal{E}_{k}^{e_{i}}=\left\lbrace e_{i}\in\mathbb{R}^{5}|e_{i}^{\top}(k)P^{\zeta}e_{i}(k)\leq\alpha_{k}^{e_{i}}\right\rbrace $, such that $\mathcal{R}_{k}^{e_{i}}\subseteq\mathcal{E}_{k}^{e_{i}}$, where the sequence $\alpha_{k}^{e_{i}}=a^{k-1}e_{i}(1)^{\top}P^{e}e_{i}(1)+(3-a)(1-a^{k-1})/(1-a)$, $P^{e}$. Since $a\in(0,1)$, {we have} that $0<1-a^{k-1}<1$ and 
				\begin{equation}
					\alpha_{k}^{e_{i}}\leq a^{k-1}e_{i}(1)^{\top}P^{e}e_{i}(1)+\alpha_{\infty}^{e_{i}},
				\end{equation}
				also because $a\in(0,1)$, for any $\epsilon>0$, there exists $k^{*}$ such that
				$
				a^{k-1}e_{i}(1)^{\top}P^{*}e_{i}(1)<\epsilon.
				$
				Therefore, for any $\epsilon>0$, there exists $k^{*}$ such that
				$
				\alpha_{k}^{e_{i}}\leq\alpha_{\infty}^{e_{i}}+\epsilon
				$, the trajectories of estimation error dynamics \eqref{er} satisfy $e_{i}(k)^{\top}P^{e}e_{i}(k)\leq\alpha_{\infty}^{e_{i}}+\epsilon$ for all $k\geq k^{*}$. By the $S$-procedure \cite{boyd2004convex}, if there exist $\tau_{1},\tau_{2}>0$ such that
				\begin{equation}\label{a3}
					\begin{split}
						&(C_{e}e_{i}(k)+\omega_{ei}(k))^{\top}\Pi(C_{e}e_{i}(k)+\omega_{ei}(k))-1\\
						&-\tau_{1}(e_{i}(k)^{\top}P^{e}e_{i}(k)-\alpha_{\infty}^{e_{i}}-\epsilon)\\
						&-\tau_{2}(\omega_{ei}(k)^{\top}\omega_{ei}(k)-\bar{\omega}_{3})\leq 0,
					\end{split}
				\end{equation}
				then, $(C_{e}e_{i}(k)+\omega_{ei}(k))^{\top}\Pi(C_{e}e_{i}(k)+\omega_{ei}(k))\leq 1$ is satisfied for all $e_{i}(k)$ and $\omega_{ei}(k)$ satisfying $e_{i}(k)^{\top}P^{e}e_{i}(k)\leq\alpha_{\infty}^{e_{i}}+\epsilon$ and $\omega_{ei}^{\top}(k)\omega_{ei}(k)\leq\bar{\omega}_{3}$. Inequality \eqref{a3} can be written as
				\begin{equation}
					\sigma(k)^{\top}\begin{bmatrix}
						f_{1}&-C_{e}^{\top}\Pi&\mathbf{0}\\
						*&\tau_{2}I-\Pi&\mathbf{0}\\
						*&*&f_{2}
					\end{bmatrix}\sigma(k)\geq 0,
				\end{equation}
				with $\sigma(k)=\begin{bmatrix}
					e_{i}(k)^{\top},\omega_{ei}(k),1
				\end{bmatrix}^{\top}$. The above inequality is satisfied if and only if $\mathcal{L}_{2}$ in \eqref{lmi3} is positive semi-definite. Therefore, for $k\geq k^{*}$, $r_{i}(k)^{\top}\Pi r_{i}(k)\leq 1$ for any $\Pi$ solution of \eqref{lmi3}. We minimize $-
				\alpha_{1}\log\det[P]-\alpha_{2}\log\det[\Pi]$ to ensure that the ellipsoidal bounds are as tight as possible.\hfill$\blacksquare$
				
				\vspace{2mm}
				Moreover, the observer gain $L$ and the monitor $\Pi$ should be chosen such that $z_{i}(k)\leq 1$ contains all attack-free trajectories of dynamics \eqref{err}, i.e., when $\delta_{i}(k)=\mathbf{0}$, there exists some $\bar{k}^{*}\in\mathbb{N}$ such that the monitor matrix $\Pi$ satisfies that $r_{i}(k)^{\top}\Pi r_{i}(k)\leq 1$ for all $k\geq \bar{k}^{*}$.

				In the attack-free case, the reachable set of the estimation error with dynamics \eqref{err} is given as
				\begin{equation}\label{re2}
					\bar{\mathcal{R}}_{k}^{e_{i}} := \left\{ \bar{e}_{i} \in \mathbb{R}^{5} \left|
					\begin{split}
						& \bar{e}_{i} \hspace{1mm}\text{satisfies \eqref{err}},\\
						&\omega_{ui}(k)^{\top}\omega_{ui}(k)\leq\bar{\omega}_{2},\\ &\omega_{ei}(k)^{\top}\omega_{ei}(k)\leq\bar{\omega}_{3},
					\end{split}
					\right. \right\}.
				\end{equation}
				
				\begin{proposition}\label{prop1}
					Consider the system matrices $(A, B_{2}, C_{e})$ and the perturbation bounds $\bar{\omega}_{2}, \bar{\omega}_{3}>0$. Assume no attacks occur {on} the system, i.e., $\delta_{i}(k)=\mathbf{0}$. For a given $c\in(0,1)$, constant $\bar{\alpha}_{\infty}^{e_{i}}:=(2-c)/(1-c)$, and $\epsilon>0$, if there exist constants $c_{1},c_{2}\in(0,1)$ and matrices $Y\in\mathbb{R}^{5\times 4}$, $P^{e}>0$ such that:
					\begin{equation}\label{lmi4}
						\left\{\begin{split}
							&c_{1},c_{2},\in(0,1), c_{1}+c_{2}\geq c, P^{e}>0, {\Pi}>0,\\
							& L_{3}:=\begin{bmatrix}
								cP^{e}&A^{\top}P^{e}-C_{e}^{\top}L^{\top}P^{e}&\mathbf{0}&\mathbf{0}\\
								*&P^{e}&-P^{e}B_{2}&-P^{e}L\\
								*&*&\frac{1-c_{1}}{\bar{\omega}_{2}}&\mathbf{0}\\
								*&*&*&\frac{1-c_{2}}{\bar{\omega}_{3}}I
							\end{bmatrix}\geq 0,\\
							& L_{4}:=\begin{bmatrix}
								\frac{1}{\bar{\alpha}_{\infty}^{e_{i}}+\epsilon+\bar{\omega}_{3}}P^{e}-C_{e}^{\top}\Pi C_{e}&-C_{e}^{\top}\Pi\\ -\Pi C_{e}&\frac{1}{\bar{\alpha}_{\infty}^{e_{i}}+\epsilon+\bar{\omega}_{3}}I-\Pi
							\end{bmatrix}\geq 0,
						\end{split}\right.
					\end{equation}
					then, the residual dynamics \eqref{err} satisfies $r_{i}^{\top}\Pi r_{i}(k)\leq 1$ for all $k\geq \bar{k}^{*}$, where \[\bar{k}^{*}:=\min\left\lbrace k\in\mathbb{N}|c^{k-1}(\bar{e}_{i}(1)^{\top}P^{e}\bar{e}_{i}(1)-\bar{\alpha}_{\infty}^{e_{i}})\leq\epsilon\right\rbrace. \]
				\end{proposition}
				\textit{Proof:} The inequality $\mathcal{L}_{3}$ is of the form \eqref{lmi1} in Corollary \ref{cor1} with $\mathcal{A}=A-LC_{e}$, $N=2$, $\mathcal{B}_{1}=-B_{2}$, $\mathcal{B}_{2}=-L$, $W_{1}=(1/\bar{\omega}_{2})$, $W_{2}=\bar{\omega}_{3}I$, $p_{1}=1$, $p_{2}=4$. From Corollary \ref{cor1}, we have $\bar{e}_{i}(k)^{\top}P^{e}e_{i}(k)\leq\bar{\alpha}_{k}^{e_{i}}$ for all $k\geq 0$, and $\bar{e}_{i}(k)$ be the solution of \eqref{ssee1}, $\bar{\alpha}_{k}^{e_{i}}=c^{k-1}\bar{e}_{i}(1)^{\top}P^{e}\bar{e}_{i}(1)+\bar{\alpha}_{\infty}^{e_{i}}(1-c^{k-1})$, and $\bar{\alpha}_{\infty}^{e_{i}}=(2-c)/(1-c)$. Since $c\in(0,1)$, for any given $\epsilon>0$, there exists $\bar{k}^{*}$ such that $\bar{e}_{i}(k)^{\top}P^{e}\bar{e}_{i}(k)\leq\bar{\alpha}_{\infty}^{e_{i}}+\epsilon$. Since $\omega_{ei}(k)^{\top}\omega_{ei}(k)\leq\bar{\omega}_{3}$, by defining $w(k):=\begin{bmatrix}
					\bar{e}_{i}(k)^{\top}&\omega_{ei}(k)^{\top}
				\end{bmatrix}^{\top}$, we have $w(k)^{\top}Q_{1}w(k)\leq 1$, with $Q_{1}=\frac{1}{\bar{\alpha}_{\infty}^{e_{i}}+\epsilon+\bar{\omega}_{3}}\diag[P^{e},I]$. Therefore, $r_{i}(k)^{\top}\Pi r_{i}(k)\leq 1$ can be written as $w(k)^{\top}Q_{2}w(k)\leq 1$ with 
				\[Q_{2}=\begin{bmatrix}
					C_{e}^{\top}\Pi C_{e}&C_{e}\Pi\\
					\Pi C_{e}&\Pi
				\end{bmatrix}.\]
				Therefore, if inequality $\mathcal{L}_{4}$ in \eqref{lmi4} is satisfied, then $Q_{2}\leq Q_{1}$, and we have $w(k)^{\top}Q_{2}w(k)\leq 1$ for all $w(k)\in\mathbb{R}^{9}$, and $k\geq\bar{k}^{*}$.
				\hfill$\blacksquare$
				
				\vspace{2mm}
				Therefore, the gains of the estimator and the detector $L$ and $\Pi$ can be obtained by solving programs \eqref{lmi3} and \eqref{lmi4} simultaneously. However, because $(L,\Pi, P^{e}, a_{3}, \tau_{1})$ are variables, the blocks $P^{e}L$, $a_{3}\Pi$, $\tau_{1}P^{e}$ are nonlinear. 
					Note that $\alpha_{\infty}^{e_{i}}+\epsilon>1$, $\tau_{1}\in(0,1)$ is a necessary condition for $f_{2}>0$, which is hence a necessary condition for $\mathcal{L}_{2}\geq 0$. To linearize the program, we let $Y=P^{e}L$ and use a grid search method over $a_{3}\in(0,1)$ and $\tau_{1}\in(0,1)$. By doing this, programs \eqref{lmi3} and \eqref{lmi4} become convex, and we can compute $L=(P^{e})^{-1}Y$.

In the following corollary of Theorem \ref{th1} and Proposition \ref{prop1}, we formulate the optimization problem for designing the observer gain $L$ and the monitor $\Pi$ such that the weighted sum of the volumes of ellipsoid $\mathcal{E}_{k}^{e_{i}}$ and $r_{i}(k)^{\top}\Pi r_{i}(k)\leq 1$ is minimized.
\begin{corollary}\label{c2}
										Consider the estimation error dynamics \eqref{er1} with matrices $(A, C_{e}, B_{2}, \Gamma)$, the perturbation bounds $\bar{\omega}_{2}$, $\bar{\omega}_{3}$. For a given $a\in(0,1)$, $c\in(0,1)$, $a_{3}\in(0,1)$, $\tau_{1}\in(0,1)$, tuning parameters $\alpha_{1},\alpha_{2}>0$ with $\alpha_{1}+\alpha_{2}=1$, $\epsilon>0$, if there exist constants $a_{1}$, $a_{2}$, $c_{1}$, $c_{2}$, $\tau_{2}$, and matrices $P^{e}$, $Y$,  $\Pi$ solution of the following optimization:
										\begin{equation}\label{lmi5}
											\left\{\begin{split}
												&\min_{P^{e},\Pi,a_{1},a_{2}, a_{3}}-\alpha_{1}\log\det[P^{e}]-\alpha_{2}\log\det[\Pi],\\
												&s.t. \hspace{1mm}a_{1},a_{2}\in(0,1), a_{1}+a_{2}+a_{3}\geq a,\\
												&c_{1}, c_{2}\in(0,1), c_{1}+c_{2}>c,\\
												& \tau_{2}\geq 0,P^{e}>0, {\Pi}>0,\\
												& L_{5}:=\begin{bmatrix}
													aP^{e}&A^{\top}P^{e}-C_{e}^{\top}(I-\Gamma\Gamma^{\dagger})^{\top}Y^{\top}&\mathbf{0}\\
													*&P^{e}&Z\\
													*&*&{W}_{a}
												\end{bmatrix}\geq 0,\\
												& L_{6}:=\begin{bmatrix}
													f_{1}&-C_{e}^{\top}\Pi&\mathbf{0}\\
													*&\tau_{2}I-\Pi&\mathbf{0}\\
													*&*&f_{2}
												\end{bmatrix}\geq 0,\\
												& L_{7}:=\begin{bmatrix}
													cP^{e}&A^{\top}P^{e}-C_{e}^{\top}Y^{\top}&\mathbf{0}&\mathbf{0}\\
													*&P^{e}&-P^{e}B_{2}&-Y\\
													*&*&\frac{1-c_{1}}{\bar{\omega}_{2}}&\mathbf{0}\\
													*&*&*&\frac{1-c_{2}}{\bar{\omega}_{3}}I
												\end{bmatrix}\geq 0,\\
												& L_{8}:=\begin{bmatrix}
													\frac{1}{\bar{\alpha}_{\infty}^{e_{i}}+\epsilon+\bar{\omega}_{3}}P^{e}-C_{e}^{\top}\Pi C_{e}&-C_{e}^{\top}\Pi\\ -\Pi C_{e}&\frac{1}{\bar{\alpha}_{\infty}^{e_{i}}+\epsilon+\bar{\omega}_{3}}I-\Pi
												\end{bmatrix}\geq 0,\\
												&f_{1}=\tau_{1}P^{e}-C_{e}^{\top}\Pi C_{e},\\
												&f_{2}=1-\tau_{1}(\alpha_{\infty}^{e_{i}}+\epsilon)-\tau_{2}\bar{\omega}_{3},
											\end{split}\right.
										\end{equation}
										with $Z=\begin{bmatrix}
											-P^{e}B_{2}&-Y(I-\Gamma\Gamma^{\dagger})&-Y\Gamma\Gamma^{\dagger}
										\end{bmatrix}$, ${W}_{a}=\diag\begin{bmatrix}
											\frac{1-a_{1}}{\bar{\omega}_{2}},\frac{1-a_{2}}{\bar{\omega}_{3}}I_{4}, (1-a_{3}){\Pi}
										\end{bmatrix}\in\mathbb{R}^{9\times 9}$, $\alpha_{\infty}^{e_{i}}=(3-a)/(1-a)$, $\bar{\alpha}_{\infty}^{e_{i}}=(2-c)/(1-c)$. Then for all $k\geq0$, $\mathcal{R}_{k}^{e_{i}}\subseteq\mathcal{E}_{k}^{e_{i}}:=\left\lbrace e_{i}\in\mathbb{R}^{5}|e_{i}^{\top}P^{e}e_{i}\leq\alpha_{k}^{e_{i}}\right\rbrace $, and $\alpha_{k}^{e_{i}}:=a^{k-1}e_{i}(1)^{\top}P^{e}e_{i}(1)+\big((3-a)(1-a^{k-1})\big)/(1-a)$. The monitor $r_{i}(k)^{\top}\Pi r_{i}(k)\leq 1$ is satisfied for all $e_{i}(k)$ satisfying \[e_{i}(k)^{\top}P^{e}e_{i}(k)\leq\alpha_{\infty}^{e_{i}}+\epsilon, \hspace{2mm}\delta_i(k)\neq\mathbf{0},\]
										and $\bar{e}_{i}(k)$ satisfying
										\[\bar{e}_{i}(k)^{\top}P^{e}\bar{e}_{i}(k)\leq\bar{\alpha}_{\infty}^{e_{i}}+\epsilon, \hspace{2mm}\delta_i(k)=\mathbf{0},\]
										and $\omega_{ei}(k)^{\top}\omega_{ei}(k)\leq\bar{\omega}_{3}$, and the weighted sum of the volumes of ellipsoid $\mathcal{E}_{k}^{e_{i}}$ and $r_{i}(k)^{\top}\Pi r_{i}(k)\leq 1$ is minimized in the sense of Corollary 1.
									\end{corollary}
									\textit{Proof:} Corollary \ref{c2} follows from Theorem \ref{th1} and Proposition \ref{prop1}.\hfill$\blacksquare$
									\begin{remark}
										By solving \eqref{lmi5}, the matrix $L$ and $\Pi$ are chosen such that the ellipsoid $r_{i}(k)^{\top}\Pi r_{i}(k)\leq 1$ contains all the possible trajectories that $\omega_{ui}(k)$, $\omega_{ei}(k)$ can induce in the residual given in \eqref{er} with $\omega_{ui}(k)^{\top}\omega_{ui}(k)\leq\bar{\omega}_{2}$, $\omega_{ei}^{\top}(k)\omega_{ei}(k)\leq\bar{\omega}_{3}$, $\delta_{i}(k)=0$. Meanwhile, the volumes of the ellipsoids $\mathcal{E}_{k}^{e_{i}}$ and $r_{i}(k)^{\top}\Pi r_{i}(k)\leq 1$ are proportional to $(\det[P^{e}])^{-1/2}$ and $(\det[\Pi])^{-1/2}$ respectively, and $(\det[P^{e}])^{-1/2}$, $(\det[\Pi])^{-1/2}$ share the same minimizer with $-\log\det[P^e]$ and $-\log\det[\Pi]$ respectively. Therefore, the objective function in \eqref{lmi5} is chosen as $-\alpha_{1}\log\det[P^{e}]-\alpha_{2}\log\det[\Pi]$ as it is convex in $P^e$ and $\Pi$ and leads to the tightest ellipsoidal bounds (in terms of volumes).
									\end{remark}
									
										\vspace{3mm}
										
										\subsection{Resilient Controller}
										We design the controller such that the attacker's reachable set is constrained to the best extent. We use Corollary 1 to obtain outer approximations of the form $\mathcal{E}_{k}^{\zeta_{i}}=\left\lbrace \zeta_{i}\in\mathbb{R}^{6}|\zeta_{i}^{\top}P^{\zeta}\zeta_{i}\leq\alpha_{k}^{\zeta_{i}}\right\rbrace $ such that $\mathcal{R}_{k}^{\zeta_{i}}\subseteq\mathcal{E}_{k}^{\zeta_{i}}$. Then, we select the controller gain $K$ such that the volume of $\mathcal{E}_{k}^{\zeta_{i}}$ is minimized. Meanwhile, matrix $K$ should be selected to guarantee the vehicle platoon attack-free performance, e.g., vehicle following and string stability objectives should be fulfilled.
										\begin{theorem}\label{th2}
											Consider the closed-loop dynamics \eqref{zeta}-\eqref{matrices} with system matrices $(A,B_{1},B_{2},a_{u},b_{u},C,C_{e})$, estimator gain $L$, monitor matrix $\Pi$, and perturbation bounds $\bar{u}, \bar{\omega}_{1},\bar{\omega}_{2},\bar{\omega}_{3}$. For a given $a\in(0,1)$, $\epsilon>0$, if there exist constants $a_{1}$, $a_{2}$, $a_{3}$, $a_{4}$, $a_{5}$, $a_{6}$ and matrices $P^{\zeta}$ and $K$ solution of the following program:
											\begin{equation}\label{lmi}
												\left\{\begin{split}
													&\min_{P^{\zeta},a_{1},\ldots, a_{6}}-\log\det[P^{\zeta}],\\
													&s.t. \hspace{1mm}a_{1},a_{2}, a_{3}, a_{4}, a_{5}, a_{6}\in(0,1),\\ &a_{1}+a_{2}+a_{3}+ a_{4}+a_{5}+a_{6}\geq a,\\
													&P^{\zeta}>0,\\
													&\mathcal{L}:\begin{bmatrix}
														aP^{\zeta}&\mathcal{A}^{\top}P^{\zeta}&\mathbf{0}\\
														P^{\zeta}\mathcal{A}&P^{\zeta}&P^{\zeta}\mathcal{B}\\
														\mathbf{0}&\mathcal{B}^{\top}P^{\zeta}&W_{a}
													\end{bmatrix}\geq 0,
												\end{split}\right.
											\end{equation}
											with $W_{a}:=\diag\begin{bmatrix}
												(1-a_{1})W_{1},\ldots,(1-a_{6})W_{6}
											\end{bmatrix}\in\mathbb{R}^{17\times 17}$, $\mathcal{B}:=(\mathcal{B}_{1},\mathcal{B}_{2},\mathcal{B}_{3},\mathcal{B}_{4},\mathcal{B}_{5},\mathcal{B}_{6})\in\mathbb{R}^{6\times 17}, W_{1}=\frac{1}{\bar{u}}, W_{2}=\frac{1}{\bar{w}_{1}}I_{2}, W_{3}=\frac{1}{\bar{w}_{2}},W_{4}=\frac{1}{\bar{w}_{3}}I_{4}, W_{5}=\Pi, W_{6}=\frac{1}{\alpha_{\infty}^{e_{i}}+\epsilon}P^{e}$; then for all $k \geq k^{*}$, $\mathcal{R}_{k}^{\zeta_{i}}\subseteq\mathcal{E}_{k}^{\zeta_{i}}:=\left\lbrace\zeta_{i}\in\mathbb{R}^{6}\vline\hspace{2mm} \zeta_{i}^{\top}P^{\zeta}\zeta_{i}\leq\alpha_{k}^{\zeta_{i}}\right\rbrace $, and $\alpha_{k}^{\zeta_{i}}:=a^{k-1}\zeta_{i}(1)^{\top}P^{*}\zeta_{i}(1)+((6-a)(1-a^{k-1}))/(1-a)$, with $\zeta_{i}(1)$ the initial value of $\zeta_{i}$, and ellipsoid $\mathcal{E}_{k}^{\zeta_{i}}$ has minimum volume in the sense of Corollary 1.
										\end{theorem}
										\textit{Proof:} Consider the reachable set $\mathcal{R}_{k}^{\zeta_{i}}$ in \eqref{rzeta}, which is a LTI system driven by peak-bounded perturbations. It follows that, under the conditions stated in Theorem \ref{th2}, Corollary \ref{cor1} can be used to obtain outer ellipsoidal approximations of the form $\mathcal{E}_{k}^{\zeta_{i}}=\left\lbrace \zeta_{i}\in\mathbb{R}^{6}|\zeta_{i}^{\top}P^{\zeta}\zeta_{i}\leq\alpha_{k}^{\zeta_{i}}\right\rbrace $ such that $\mathcal{R}_{k}^{\zeta_{i}}\subseteq\mathcal{E}_{k}^{\zeta_{i}}$, where the sequence $\alpha_{k}^{\zeta_{i}}$ is given by $\alpha_{k}^{\zeta_{i}}=a^{k-1}\zeta_{i}(1)^{\top}P^{\zeta_{i}}\zeta_{i}(1)+(6-a)(1-a^{k-1})/(1-a)$, $P^{\zeta}$ is the solution of \eqref{lmi}. The volume of $\mathcal{E}_{k}^{\zeta_{i}}$ is minimal in the sense of Corollary \ref{cor1}.\hfill$\blacksquare$
		
										\vspace{2mm}
										Next, we select matrix $K$ to guarantee the controller performance in the attack-free case. First, $K=\begin{bmatrix}
											k_{p}&k_{d}
										\end{bmatrix}$ should satisfy that $k_{p},k_{d}>0$ with $k_{d}>k_{p}\tau$ in order to fulfill the vehicle following and string stability objectives \cite{Ploeg2014}. Besides, the controller needs to ensure that the vehicle error dynamics has an acceptable decay rate:
										\begin{equation}\label{error}
											\begin{split}
												\begin{bmatrix}
													\dot{e}_{1,i}\\
													\dot{e}_{2,i}\\
													\dot{e}_{3,i}
												\end{bmatrix}=\begin{bmatrix}
													0&1&0\\
													0&0&1\\
													0&0&-\frac{1}{\tau}
												\end{bmatrix}\begin{bmatrix}
													e_{1,i}\\
													e_{2,i}\\
													e_{3,i}
												\end{bmatrix}+\begin{bmatrix}
													0\\0\\-\frac{1}{\tau}
												\end{bmatrix}u_{i}, \hspace{2mm}2\leq i\leq m,
											\end{split}
										\end{equation}
										with $u_{i}=K\begin{bmatrix}
											e_{1,i}\\
											e_{2,i}
										\end{bmatrix}$, $K=\begin{bmatrix}
											k_{p}&k_{d}
										\end{bmatrix}$, and
										\begin{equation}
											\begin{bmatrix}
												e_{1,i}\\
												e_{2,i}\\
												e_{3,i}
											\end{bmatrix}=\begin{bmatrix}
												e_{ri}\\
												\dot{e}_{ri}\\
												\ddot{e}_{ri}
											\end{bmatrix},\hspace{2mm}2\leq i\leq m.
										\end{equation}
										Substituting $u_{i}=\begin{bmatrix}
											k_{p}&k_{d}
										\end{bmatrix}\begin{bmatrix}
											e_{1,i}\\
											e_{2,i}
										\end{bmatrix}$ into \eqref{error} yields
										\begin{equation}\label{error2}
											\begin{bmatrix}
												\dot{e}_{1,i}\\
												\dot{e}_{2,i}\\
												\dot{e}_{3,i}
											\end{bmatrix}=\underbrace{\begin{bmatrix}
													0&1&0\\
													0&0&1\\
													-\frac{k_{p}}{\tau}&-\frac{k_{d}}{\tau}&-\frac{1}{\tau}
											\end{bmatrix}}_{A_{e}}		\begin{bmatrix}
												e_{1,i}\\
												e_{2,i}\\
												e_{3,i}
											\end{bmatrix}.
										\end{equation}
										We enforce the decay rate of system \eqref{error2} by enforcing the real parts of all the eigenvalues of $A_{e}$ to be smaller than a given constant $\lambda_{\max}<0$. This can be achieved by letting 
										\begin{equation}
											\left\{\begin{split}
												-\frac{1}{3\tau}<&\lambda_{\max},\\
												\frac{1}{3\tau}>k_{d}>&-2\lambda_{\max}-3\tau\lambda_{\max}^{2},\\
												k_{p}>&-\lambda_{\max}k_{d}-\lambda_{\max}^2-\tau\lambda_{\max}^{3}.
											\end{split}\right.
										\end{equation}
											
											In the following corollary, we formulate the optimization problem for designing the controller $K$ such that the volume of ellipsoid $\mathcal{E}_{k}^{\zeta_{i}}$ is minimized and vehicle following and string stability objectives in the attack-free case are fulfilled with decay rate of system \eqref{error} enforced.
											
											\begin{corollary}\label{cor3}
												Consider the closed-loop dynamics \eqref{zeta}-\eqref{matrices} with system matrices $(A,B_{1},B_{2},a_{u},b_{u},C,C_{e})$, estimator gain $L$, monitor matrix $\Pi$, and perturbation bounds $\bar{u}, \bar{\omega}_{1},\bar{\omega}_{2},\bar{\omega}_{3}$. For a given $a\in(0,1)$, $\epsilon>0$, $\lambda_{\max}<0$, $\tau\in(0,-\frac{1}{3\lambda_{\max}})$, if there exist constants $a_{1}$, $a_{2}$, $a_{3}$, $a_{4}$, $a_{5}$, $a_{6}$, $\gamma>0$, and matrices $P^{\zeta}$, $P_{s}$, and $K=\begin{bmatrix}
													k_{p}&k_{d}
												\end{bmatrix}$ solution of
												\begin{equation}\label{lmi6}
													\left\{\begin{split}
														&\min_{P^{\zeta},a_{1},\ldots, a_{6}}-\log\det[P^{\zeta}],\\
														&s.t. \hspace{1mm}a_{1},a_{2}, a_{3}, a_{4}, a_{5}, a_{6}\in(0,1),\\ &a_{1}+a_{2}+a_{3}+ a_{4}+a_{5}+a_{6}\geq a,\\
														&k_{p}>0,k_{d}>0,k_{d}>k_{p}\tau,\\
														& k_{d}>-2\lambda_{\max}-3\tau\lambda_{\max}^{2},\\
														&k_{p}>-\lambda_{\max}k_{d}-\lambda_{\max}^2-\tau\lambda_{\max}^{3},\\
														&P^{\zeta}>0,\\
														&\mathcal{L}:\begin{bmatrix}
															aP^{\zeta}&\mathcal{A}^{\top}P^{\zeta}&\mathbf{0}\\
															P^{\zeta}\mathcal{A}&P^{\zeta}&P^{\zeta}\mathcal{B}\\
															\mathbf{0}&\mathcal{B}^{\top}P^{\zeta}&W_{a}
														\end{bmatrix}\geq 0,
													\end{split}\right.
												\end{equation}
												with $W_{a}:=\diag\begin{bmatrix}
													(1-a_{1})W_{1},\ldots,(1-a_{6})W_{6}
												\end{bmatrix}\in\mathbb{R}^{17\times 17}$, $\mathcal{B}:=(\mathcal{B}_{1},\mathcal{B}_{2},\mathcal{B}_{3},\mathcal{B}_{4},\mathcal{B}_{5},\mathcal{B}_{6})\in\mathbb{R}^{6\times 17}, W_{1}=\frac{1}{\bar{u}}, W_{2}=\frac{1}{\bar{w}_{1}}I_{2}, W_{3}=\frac{1}{\bar{w}_{2}},W_{4}=\frac{1}{\bar{w}_{3}}I_{4}, W_{5}=\Pi, W_{6}=\frac{1}{\alpha_{\infty}^{e_{i}}+\epsilon}P^{e}$, $\tilde{\mathcal{B}}=(\mathcal{B}_{1},\mathcal{B}_{2},\mathcal{B}_{2})\in\mathbb{R}^{6\times 4}$; then for all $k \geq k^{*}$, $\mathcal{R}_{k}^{\zeta_{i}}\subseteq\mathcal{E}_{k}^{\zeta_{i}}:=\left\lbrace \zeta_{i}^{\top}P^{\zeta}\zeta_{i}\leq\alpha_{k}^{\zeta_{i}}\right\rbrace $, and $\alpha_{k}^{\zeta_{i}}:=a^{k-1}\zeta_{i}(1)^{\top}P^{*}\zeta_{i}(1)+((6-a)(1-a^{k-1}))/(1-a)$, with $\zeta_{i}(1)$ the initial value of $\zeta_{i}$, the ellipsoid $\mathcal{E}_{k}^{\zeta_{i}}$ has minimum volume in the sense of Corollary 1; vehicle following and string stability objectives are satisfied and the real parts of all the eigenvalues of $A_{e}$ are smaller than $\lambda_{\max}$. 
											\end{corollary}
											Note that $P^{\zeta}$ and $K$ are both variables, the terms $P^{\zeta}\mathcal{A}$, $P^{\zeta}\mathcal{B}$ are nonlinear. To linearize the program \eqref{lmi6}, we impose some structure on the matrices  $P^{\zeta}$.  Let $P^{\zeta}$ be positive definite and of the form 
											\begin{equation}
												P^{\zeta}:=\begin{bmatrix}
													X&\mathbf{0}\\
													\mathbf{0}&\tilde{x}
												\end{bmatrix},
											\end{equation}
											with $X\in\mathbb{R}^{5\times 5}$, $X>0$, and $\tilde{x}>0$.
											We have 
											\begin{equation}\label{pa}
												P^{\zeta}\mathcal{A}=\begin{bmatrix}
													XA&XB_1\\
													b_{u}\tilde{x}KC&a_{u}\tilde{x}
												\end{bmatrix},
											\end{equation}
											and
											\begin{equation}\label{pb}
												\begin{split}
													&P^{\zeta}\mathcal{B}=\\
													&\begin{bmatrix}
													XB_{2}&\mathbf{0}&\mathbf{0}&\mathbf{0}&\mathbf{0}&\mathbf{0}\\
													b_{u}\tilde{x}&b_{u}\tilde{x}K&b_{u}\tilde{x}&-b_{u}\tilde{x}K\Gamma^{\dagger}&b_{u}\tilde{x}K\Gamma^{\dagger}&-b_{u}\tilde{x}K\Gamma^{\dagger}C_{e}
												\end{bmatrix}.
												\end{split}
											\end{equation}
											We define a new variable $\tilde{K}:=\tilde{x}K$ and \eqref{pa}-\eqref{pb} become linear in $X, \tilde{x}$ and $\tilde{K}$, and hence $\mathcal{L}$ in \eqref{lmi6} is linearized.

											Hence, the program \eqref{lmi} becomes convex. By solving the program in \eqref{lmi}, we can compute $K=\frac{\tilde{K}}{\tilde{x}}$ and obtain $P^{\zeta}$ and use the following corollary for obtaining $P^{x}$ and $\mathcal{E}_{k}^{x_{i}}$. 
											\begin{corollary}
												Let the conditions of Theorem \ref{th2} be satisfied and consider the corresponding matrix $P^{\zeta_{i}}$ and function $\alpha_{k}^{\zeta_{i}}$. Let $P^{\zeta_{i}}$ be partitioned as
												\begin{equation}
													P^{\zeta_{i}}=:\begin{bmatrix}
														P_{1}^{\zeta_{i}}&P_{2}^{\zeta_{i}}\\
														P_{2}^{\zeta_{i}^{\top}}&P_{3}^{\zeta_{i}}
													\end{bmatrix},
												\end{equation}
												with $P_{1}^{\zeta_{i}}\in\mathbb{R}^{5}$, $P_{2}^{\zeta_{i}}\in\mathbb{R}^{5\times 1}$, and $P_{3}^{\zeta_{i}}\in R$. Then, for $k\geq k^{*}$, $\mathcal{R}_{k}^{x_{i}}\subseteq\mathcal{E}_{k}^{x_{i}}:=\left\lbrace x_{i}\in\mathbb{R}^{5}|x_{i}^{\top}P^{x}x_{i}\leq\alpha_{k}^{x_{i}}\right\rbrace $ with $P^{x}:=P_{1}^{\zeta_{i}}-P_{2}^{\zeta_{i}}(P_{3}^{\zeta_{i}})^{-1}(P_{2}^{\zeta_{i}})^{\top}$ and $\alpha_{k}^{x_{i}}:=\alpha_{k}^{\zeta_{i}}$.
											\end{corollary}
											\textit{Proof:} By theorem \ref{th2}, the trajectories of \eqref{zeta} satisfy $\zeta_{i}^{\top}P^{\zeta}\zeta_{i}\leq\alpha_{k}^{\zeta_{i}}$ for $k\geq k^{*}$. By Lemma \ref{lemma1}, the projection of $\zeta_{i}(k)^{\top}P^{\zeta}\zeta_{i}(k)\leq\alpha_{k}^{\zeta_{i}}$ onto the $x_{i}$-hyperplane is given by $\mathcal{E}_{k}^{x_{i}}$ defined above. Thus, in light of Remark \ref{remark2}, the trajectories of the vehicle dynamics are contained in $\mathcal{E}_{k}^{x_{i}}$, i.e., $\mathcal{R}_{k}^{x_{i}}\subseteq\mathcal{E}_{k}^{x_{i}}$ for all $k\geq k^{*}$.\hfill$\blacksquare$
											\vspace{2mm}
											\begin{remark}
													Since the gains of the estimator, the monitor, and the controller, i.e., $L,\Pi,$ and $K$ are all selected with the purposed of minimizing the volumes of $\mathcal{E}_{k}^{\zeta_{i}}$, and $\mathcal{E}_{k}^{x_{i}}$ is the projection of $\mathcal{E}_{k}^{\zeta_{i}}$ onto the $x_{i}$-hyperplane, hence, the volume of $\mathcal{E}_{k}^{x_{i}}$ is also minimized. This indicates our design approaches minimize the platooning performance deterioration caused by stealthy attacks.
											\end{remark}
											\section{A Security Assessment: Distance to Critical States}\label{critical}
											In this section, we aim to provide a security assessment method for evaluating CAV resilience to cyberattacks. . We introduce the notion of \emph{critical states} $\mathcal{C}^{x_{i}}$ -- states that, if reached, compromise the integrity or safe operation of the vehicles. Such a region may represent states in which, a collision between two neighboring vehicles occurs, or the velocity of the vehicle exceeds the maximum allowed value. So far, we have characterized all the states that attacks can induce in the platoon while being stealthy to the system monitor (the stealthy reachable set $\mathcal{R}_{k}^{x_{i}}$). Therefore, if the intersection between $\mathcal{C}^{x_{i}}$ and $\mathcal{R}_{k}^{x_{i}}$ is not empty, there exist attack sequences $\delta_i$ that can induce a critical event by tampering with radars. Conversely, if such an intersection is empty, we can ensure that critical events cannot be induced by stealthy attacks on radars. Here, we compute the minimum distance $d_{k}^{x_{i}}$ from $\mathcal{E}_{k}^{x_{i}}$ to $\mathcal{C}^{x_{i}}$ and use it as an approximation of the distance between $\mathcal{R}_{k}^{x_{i}}$ and $\mathcal{C}^{x_{i}}$ (because computing $\mathcal{R}_{k}^{x_{i}}$ exactly is not tractable, we only have the approximation $\mathcal{E}_{k}^{x_{i}}$). The distance  $d_{k}^{x_{i}}$ gives us intuition of how far the actual reachable set  $\mathcal{R}_{k}^{x_{i}}$ is from $\mathcal{C}^{x_{i}}$.
											
\begin{figure}[t]\centering
												\includegraphics[width=0.4\textwidth]{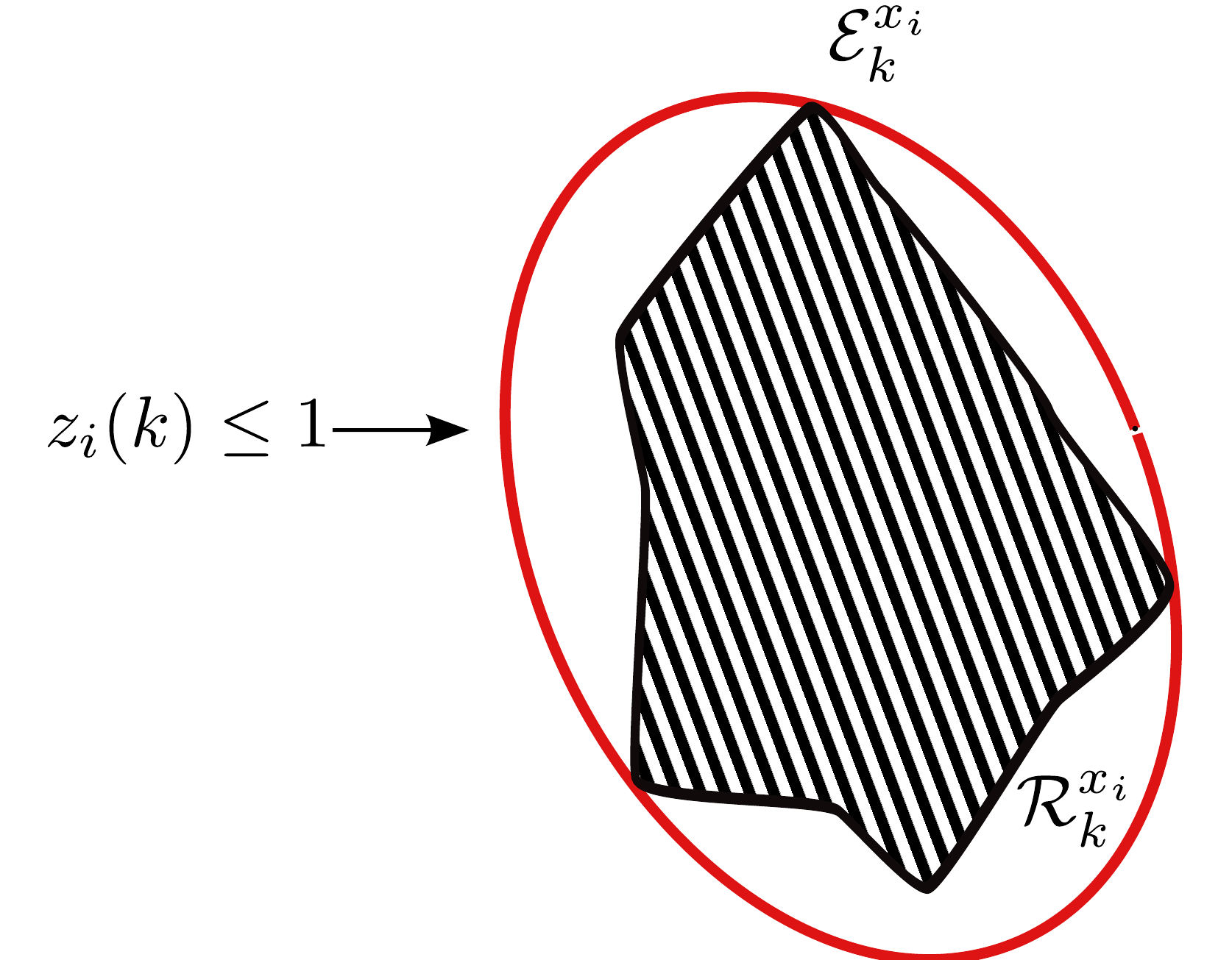}
												\caption{Stealthy reachable set of the $i$-th CAV $\mathcal{R}_{k}^{x_{i}}$ and outer ellipsoidal approximation $\mathcal{E}_{k}^{x_{i}}$ of $\mathcal{R}_{k}^{x_{i}}$.}
												\centering
												\label{fig:2s}
											\end{figure}

\begin{figure}[t]\centering
												\includegraphics[width=0.4\textwidth]{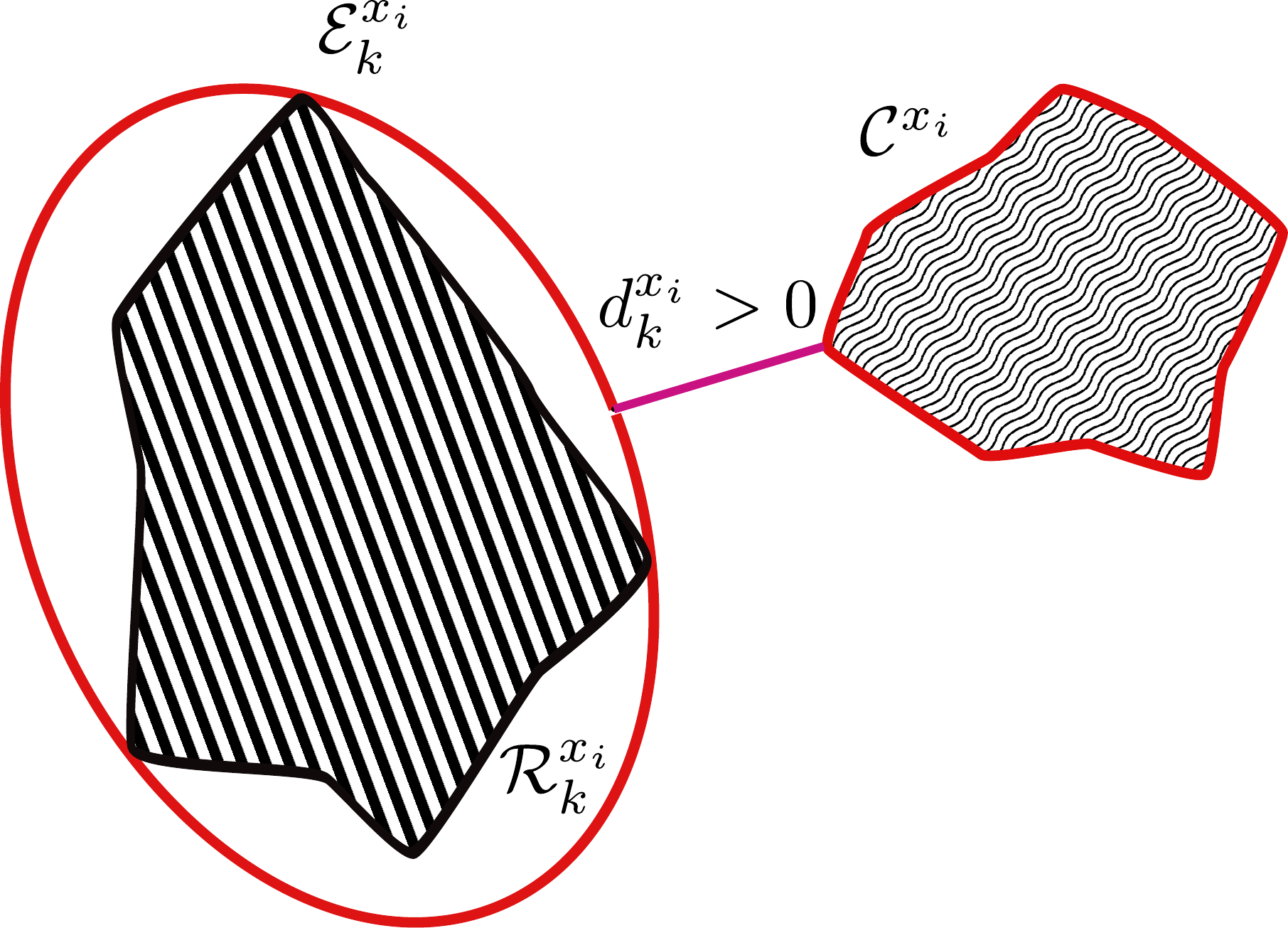}
												\caption{Minimum distance $d_{k}^{x_{i}}>0$ implies the $i$-th CAV is resilient to stealthy FDI attacks.}
												\centering
												\label{fig:3s}
											\end{figure}

\begin{figure}[t]\centering
												\includegraphics[width=0.4\textwidth]{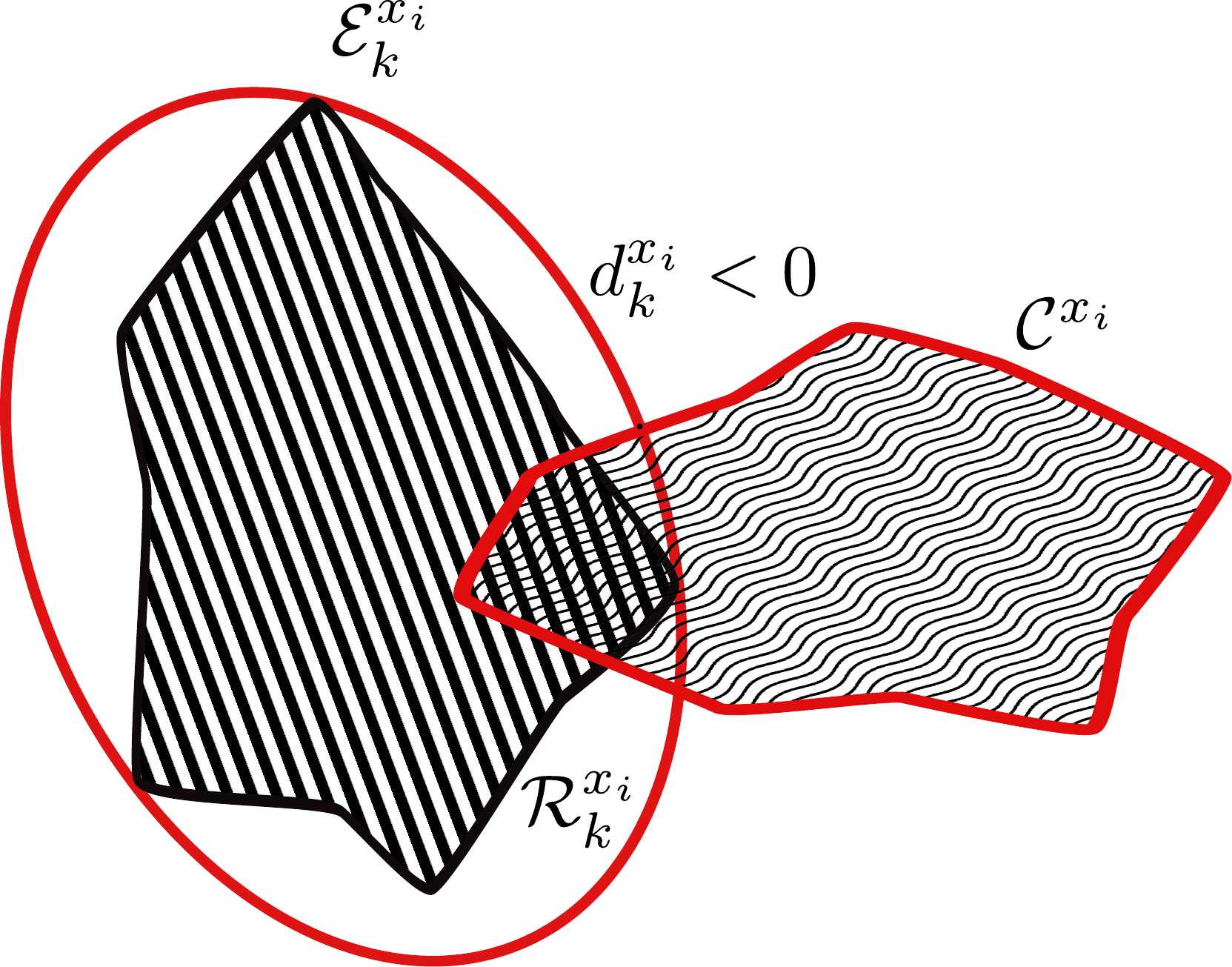}
												\caption{Minimum distance $d_{k}^{x_{i}}<0$ and there is an intersection between $\mathcal{R}_{k}^{x_{i}}$ and $\mathcal{C}^{x_{i}}$ -  the $i$-th CAV is vulnerable to stealthy FDI attacks.}
												\centering
												\label{fig:4s}
											\end{figure}

\begin{figure}[t]\centering
												\includegraphics[width=0.4\textwidth]{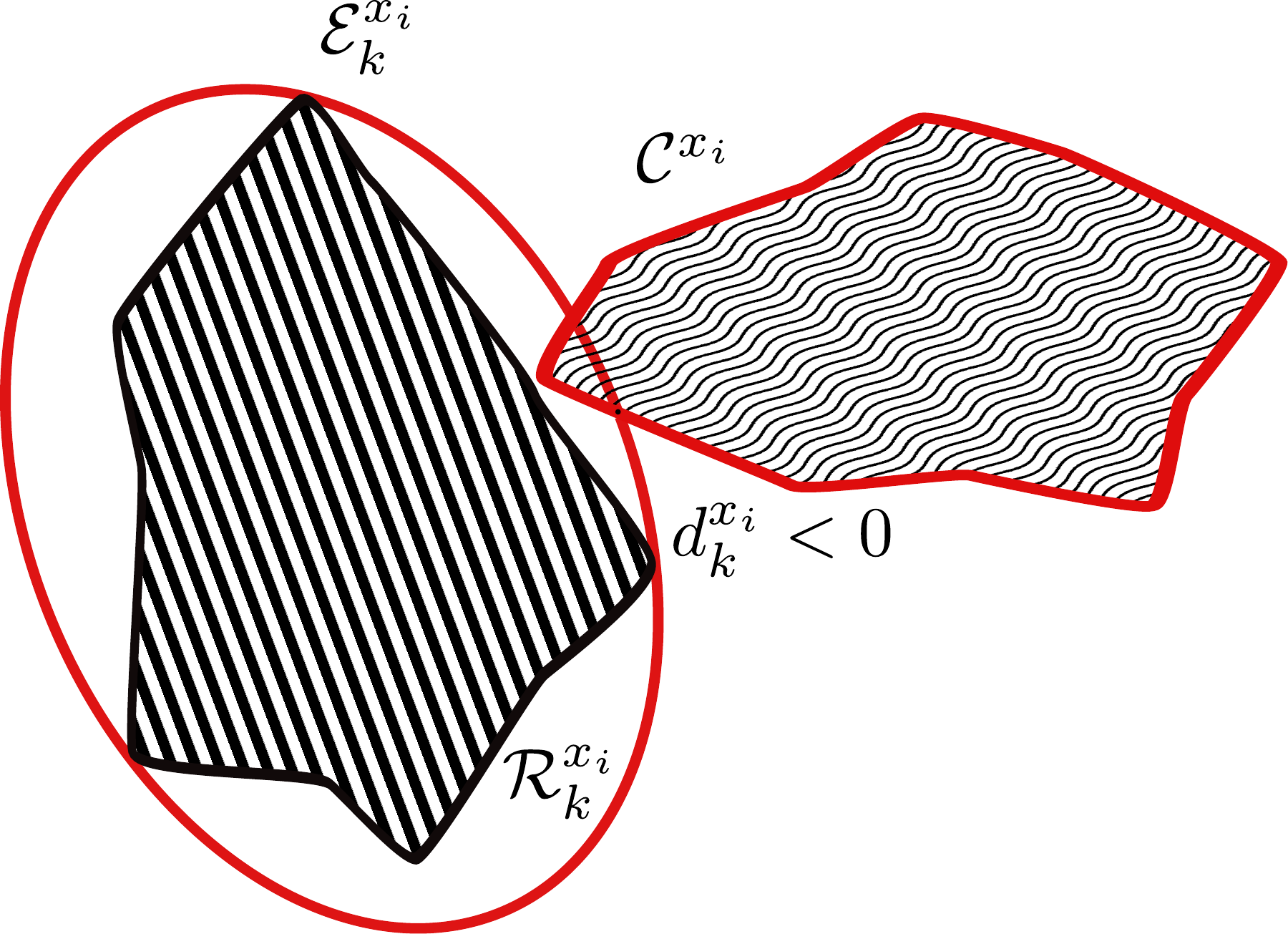}
												\caption{Minimum distance $d_{k}^{x_{i}}<0$ but there is no intersection between $\mathcal{R}_{k}^{x_{i}}$ and $\mathcal{C}^{x_{i}}$ -  the $i$-th CAV is resilient to stealthy FDI attacks.}
												\centering
												\label{fig:5s}
											\end{figure}

											From \eqref{e}, we have that the tracking error of each vehicle at the sampling time-instants is given by
											\begin{equation}
												e_{ri}(k)=d_{i}(k)-(s_{i}+hv_{i}(k)).
											\end{equation}
											Since $d_{i}(k)< 0$ indicates a collision between vehicles $i$ and $i-1$, one subset of critical states is clearly given by:
											\begin{equation}\label{cc1}
												\begin{split}
													e_{ri}(k)<-s_{i}-hv_{i}(k).
												\end{split}
											\end{equation}
											Another potential subset of critical states is that the maximum allowed velocity $v_{\max}$ is exceeded, i.e.,
											\begin{equation}\label{cc2}
												v_{i}(k) > v_{\max}.
											\end{equation}
											Combining \eqref{cc1} and \eqref{cc2}, the set of critical states is characterized by the union of half-spaces:
											\begin{equation}\label{c}
												\mathcal{C}^{x_{i}}:=\big\lbrace x_{i}\in\mathbb{R}^{5}\vline \hspace{1mm}c_{1}^{\top}x_{i}> s_{i} \text{ and } c_{2}^{\top}x_{i}> v_{\max}\big\rbrace,
											\end{equation}
											with $c_{1}^{\top}=\begin{bmatrix}
												-1&-h&0&0&0
											\end{bmatrix}$, and $c_{2}^{\top}=\begin{bmatrix}
												0&1&0&0&0
											\end{bmatrix}$.
											\begin{corollary}
												Consider the set of critical states $\mathcal{C}^{x_{i}}$ defined in \eqref{c} and the matrix $P^{x}$ and the function $\alpha_{k}^{x_{i}}$ obtained in Theorem 1. The minimum distance, $d_{k}^{x_{i}}$, between the outer ellipsoidal approximation of $\mathcal{R}_{k}^{x_{i}}$, $\mathcal{E}_{k}^{x_{i}}=\left\lbrace x_{i}\in\mathbb{R}^{5}\vline \hspace{1mm}x_{i}^{\top}P^{x}x_{i}\leq\alpha_{k}^{x_{i}}\right\rbrace $, and $\mathcal{C}^{x_{i}}$ is given by
												\begin{equation}\label{dd}
													d_{k}^{x_{i}}=\min\left(\frac{|b_{j}|-\sqrt{c_{j}^{\top}(P^{x})^{-1}c_{j}/\alpha_{k}^{x_{i}}}}{c_{j}^{\top}c_{j}} \right) , j=1,2.
												\end{equation}
											\end{corollary}
											\textit{Proof:} The minimum distance between an ellipsoid centered at the origin $\left\lbrace x\in\mathbb{R}^{n}|x^{\top}Px=1\right\rbrace $, $P>0$, and a hyperplane $\left\lbrace x\in\mathbb{R}^{n}|\hspace{1mm}c^{\top}x=b\right\rbrace $, $c\in\mathbb{R}^{n}$, $b\in\mathbb{R}$ is given by $(|b|-\sqrt{c^{\top}P^{-1}c})/c^{\top}c$ \cite{kurzhanskiy2006ellipsoidal}. The minimum distance between $\mathcal{C}^{x_{i}}$ and $\mathcal{E}^{x_{i}}_{k}$ is hence given by \eqref{dd}.\hfill$\blacksquare$
											\begin{remark}
												If $d_{k}^{x_{i}}> 0$ for all $k \in \mathbb{N}$, the ellipsoid $\mathcal{E}_{k}^{x_{i}}$ does not intersect with the set of critical states $\mathcal{C}^{x_{i}}$. Since $\mathcal{R}_{k}^{x_{i}}\subseteq\mathcal{E}_{k}^{x_{i}}$, the intersection between $\mathcal{C}^{x_{i}}$ and $\mathcal{R}_{k}^{x_{i}}$ is also empty. This indicates no collision will occur between the $i$-th and the $i-1$-th CAVs and its velocity will not exceed $v_{\max}$ under stealthy attacks. We can thus conclude the $i$-th CAV is resilient to stealthy attacks on radars. A schematic representation of this idea is given in Figures \ref{fig:2s}-\ref{fig:3s}. On the contrary, if $d_{k}^{x_{i}}< 0$ for some $k \in \mathbb{N}$, the intersection between the ellipsoid $\mathcal{E}_{k}^{x_{i}}$ and the set of critical states $\mathcal{C}^{x_{i}}$ is not empty. This indicates that collisions between vehicles $i$ and $i-1$ is possible or the $i$-th CAV might be speeding under stealthy attacks. The latter case is depicted in Figure \ref{fig:4s}. However, due to the potential conservatism of the ellipsoidal bounds, the scenario depicted in Figure \ref{fig:5s} may also occur, i.e., $d_{k}^{x_{i}}<0$ but the intersection between $\mathcal{R}_{k}^{x_{i}}$ and $\mathcal{C}^{x_{i}}$ is empty. Therefore, if $d_{k}^{x_{i}}> 0$, the $i$-th CAV is risk-free in the presence of stealthy FDI attacks; otherwise, the $i$-th CAV is under risk.
											\end{remark}
											
											\section{Simulation}\label{simulation}
											In this section, we use a simulation study to illustrate the effectiveness of our approaches.
											
\begin{figure}[t]\centering
												\includegraphics[width=0.53\textwidth]{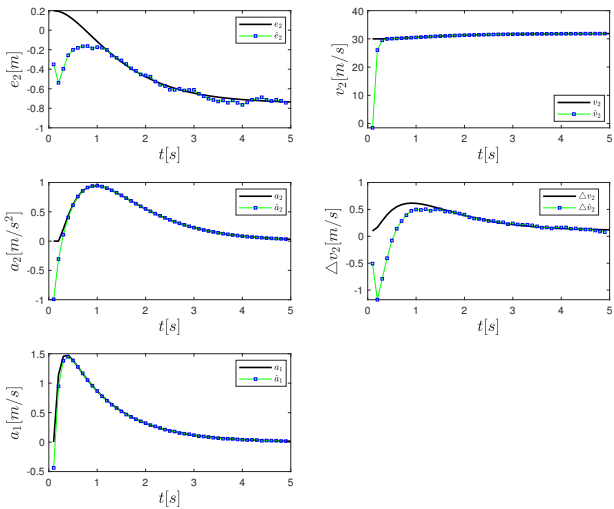}
												\caption{The estimation $\hat{x}_{2}(k)$ converges to the true vehicle state $x_{2}(k).$}
												\centering
												\label{fig:2}
											\end{figure}

											Consider two homogeneous vehicles in a platoon with initial speed $30m/s$. Suppose $h=0.5$, $\tau=0.1$. Using exact discretization approach, we obtain the discrete-time system \eqref{closed} with sampling interval $T_{s}=0.1$ seconds. Suppose the desired acceleration of vehicle $i$ ranges from $-2m/s^2$ to $2m/s^2$, i.e., $u_{1}\in[-2,2]$. At each time step $k$, $u_{1}(k)$ is transmitted from vehicle $1$ to vehicle $2$. {Disturbances $\omega_{d2}$, $\omega_{u2}$ are uniformly distributed in the intervals $(-0.1,0.1)$ and $(-0.01,0.01)$ respectively, i.e., $\omega_{d2}\in\mathcal{U}(-0.1,0.1)$, $\omega_{u2}\in\mathcal{U}(-0.01,0.01)$.} Let $\bar{u}=4$, $\bar{\omega}_{1}=0.01$, $\bar{\omega}_{2}=0.0001$, $\bar{\omega}_{3}=0.02$. At each time step $k$, $u_{1}(k)$ is transmitted from vehicle $1$ to vehicle $2$ via the communication network. We use Corollary \ref{c2} to obtain the optimal state estimator and the monitor matrix $Pi$. We let the tuning parameters $\alpha_{1}=0.95$, $\alpha_{2}=0.05$, by solving \eqref{lmi5} using a grid search over $a\in(0,1)$, $c\in(0,1)$, $a_{3}\in(0,1)$, and $\tau_{1}\in(0,1)$, we can obtain the estimator and monitor matrices as follows:
											\begin{equation}
												L=\begin{bmatrix}\label{l}
													0.2773  & -0.0109  & -0.0382  &  0.0000\\
													-0.0000  &  0.8754  &  0.0118  & -0.0000\\
													-0.0000  &  0.0006  &  0.0403  & -0.0000\\
													0.0000 &  -0.0265  & -0.0961  &  0.2772\\
													-0.0001  & -0.0007  & -0.0123  &  0.0013
												\end{bmatrix},
											\end{equation}
											\begin{equation}\label{pi}
												\Pi=\begin{bmatrix}
													0.0001  &  0.0000 &  -0.0000 &   0.0000\\
													0.0000  &  0.2470 &  -0.0001  &  0.0000\\
													-0.0000  & -0.0001  &  0.2578 &  -0.0000\\
													0.0000  &  0.0000 &  -0.0000  &  0.0001
												\end{bmatrix}.
											\end{equation}
											Next, we use Corollary \ref{cor3} to obtain the optimal controller. We solve \eqref{lmi6} and enforce the real parts of all the eigenvalues of $A_{e}$ to be smaller than $\lambda_{\max}=-0.01$ and we have 
											\begin{equation}
												K=\begin{bmatrix}
													0.0051 & 0.0204
												\end{bmatrix}.
											\end{equation}
											Suppose $u_{1}(k)=2e^{-0.1k}$, the initial condition of vehicle $2$ is given as $x_{2}(1)=\begin{bmatrix}
												0&30&0&0&0
											\end{bmatrix}$. The initial condition of the estimator $\hat{x}_{2}(1)$ is randomly chosen. The performance of the estimator is shown in Figure \ref{fig:2}.
											Suppose the maximum allowed speed is $35m/s$, i.e., $v_{\max}=35$; the average standstill distance $s_{2}=3$ meters is adopted as given in \cite{houchin2015measurement}. Therefore, In Section \ref{critical}, the ellipsoidal approximation $\mathcal{E}_{k}^{x_{i}}$ is $k$-dependent; however, since $a\in(0,1)$,  the function $\alpha_{k}^{x_{i}}$ conforming $\mathcal{E}_{k}^{x_{i}}$ converges to $\alpha_{\infty}^{x_{i}}=(6-a)/(1-a)$ exponentially. Therefore, in a few time steps, $\mathcal{E}_{k}^{x_{2}}\approx\mathcal{E}_{\infty}^{x_{2}}=\left\lbrace x_{2}\in\mathbb{R}^{5}|\hspace{2mm}x_{2}^{\top}P^{x}x_{2}\leq (6-a)/(1-a)\right\rbrace $. Therefore, we compute the distance between $C^{x_{2}}$ and $\mathcal{E}_{\infty}^{x_{2}}$ as
											\begin{equation}\label{ddd}
												d_{\infty}^{x_{2}}=\min\left(\frac{|b_{j}|-\sqrt{c_{j}^{\top}(P^{x})^{-1}c_{j}/\alpha_{\infty}^{x_{2}}}}{c_{j}^{\top}c_{j}} \right) , j=1,2.
											\end{equation}
											We obtain that $d_{\infty}^{x_{2}}=1.3978$, which is positive. This indicates that the vehicle equipped with estimator $L$ and monitor $\Pi$ given as \eqref{l} and \eqref{pi} is resilient to stealthy attacks on radars. However, if we design estimator $L$, monitor $\Pi$ and controller $K$ without using our synthesis approach, $d_{\infty}^{x_{2}}$ can be negative, which indicates the vehicle platoon is under risk in the presence of stealthy attacks on radars. For example, if we consider designing an estimator optimal in terms of minimizing the effect of disturbances on estimation error, we can choose $L$ such that the $H_{\infty}$ gain from $\omega_{2}(k)=\begin{bmatrix}
												\omega_{u2}(k)&\omega_{e2}^{\top}(k)
											\end{bmatrix}^{\top}$ to $e_{2}(k)$ is minimized, we can obtain the estimator gain given as follows:
											\begin{equation}\label{ll}
												\tilde{L}=\begin{bmatrix}
													0.0245 &  -0.0001   & 0.0001 &  -0.0994\\
													-0.0001  &  0.0147  & -0.0001 &  -0.0000\\
													0.0000  & -0.0001  &  0.0012  &  0.0000\\
													-0.0994  & -0.0000  & -0.0000 &   1.0083\\
													-0.0105  & -0.0000  &  0.0001 &   0.1053
												\end{bmatrix}
											\end{equation}
											with $H_{\infty}$ gain equal to $1.0425$ \cite{dullerud2013course}. Solving \eqref{lmi4} by substituting $L$ as given in \eqref{ll}, we can obtain the monitor with $\tilde{\Pi}$ given as:
											\begin{equation}
												\tilde{\Pi}=\begin{bmatrix}
													0.1863  &  0.0001  & -0.0002  &  0.0173\\
													0.0001  &  0.1879 &   0.0004 &   0.0000\\
													-0.0002 &   0.0004  &  0.2239 &  -0.0000\\
													0.0173  &  0.0000  & -0.0000  &  0.0151
												\end{bmatrix}.
											\end{equation}
											Moreover, suppose we adopt the controller introduced in \cite{Ploeg2014} with $\tilde{K}=\begin{bmatrix}
												0.2&0.7
											\end{bmatrix}$. Note that such a controller gain is selected in \cite{Ploeg2014} to fulfill the vehicle following and string stability objectives without security considerations. We can now compute $d_{\infty}^{x_{2}}$ from \eqref{ddd} and obtain $d_{\infty}^{x_{2}}=-3.738\times 10^{3}$. This indicates that vehicle $2$ equipped with estimator $\tilde{L}$, monitor $\tilde{\Pi}$, and controller $\tilde{K}$ might collide with vehicle $1$ or might be speeding if its radars are under stealthy attacks.
											\section{Conclusion}\label{conclusion}
											We have provided algorithmic tools for designing estimators, fault/attack detectors and controllers for enhancing the resilience of CAVs in a platoon to sensor attacks. We have given design methods for constructing the estimator, the detector and the controller such that their attack-free performances are guaranteed, i.e., the estimator and detector can be used for state estimation and identifying faults or general FDI attacks on CAV sensors and the controller can fulfill vehicle following and string stability objectives; and at the same time, they minimize the degradation of platooning performance caused by stealthy sensor attacks -- attacks that are intelligent enough to surpass the detector by hiding within the system uncertainties. A security assessment method has been provided to evaluate the resilience of CAVs to stealthy attacks. The method has also been used to show the effectiveness of our approaches. 
																											\bibliographystyle{ieeetr}
																											\bibliography{Observer1}
																										\end{document}